\documentclass[%
 aip,jcp,floatfix,
 amsmath,amssymb,
 reprint,%
]{revtex4-1}

\usepackage[dvips]{graphicx}
\usepackage{bm}% bold math
\usepackage[dvipsnames]{xcolor}
\usepackage{soul}
\usepackage{amsmath}
\usepackage{amssymb}
\usepackage[normalem]{ulem}
\usepackage[utf8]{inputenc}
\usepackage[T1]{fontenc}
\usepackage{mathptmx}
\usepackage{etoolbox}
\usepackage{hyperref}
\usepackage{natbib}
\usepackage[mathlines]{lineno}% Enable numbering of text and display math
\usepackage{placeins}
\usepackage[caption=false,font=normalsize,labelfont=bf]{subfig}

\makeatletter
\def\@email#1#2{%
 \endgroup
 \patchcmd{\titleblock@produce}
  {\frontmatter@RRAPformat}
  {\frontmatter@RRAPformat{\produce@RRAP{*#1\href{mailto:#2}{#2}}}\frontmatter@RRAPformat}
  {}{}
}%
\makeatother
\begin{document}

%\preprint{AIP/123-QED}

\title[]{Kinetic reconstruction of free energies as a function of multiple order parameters}
% Force line breaks with \\
\author{Yagyik Goswami}
\author{Srikanth Sastry}%
 \email{sastry@jncasr.ac.in}
 \homepage{https://www.jncasr.ac.in/faculty/sastry/}
\affiliation{ 
Theoretical Sciences Unit and School of Advanced Materials, Jawaharlal Nehru Centre for Advanced Scientific Research, Bengaluru, India.%\\This line break forced with \textbackslash\textbackslash
}%

% \homepage{http://www.Second.institution.edu/~Charlie.Author.}
%\affiliation{%
%Second institution and/or address%\\This line break forced% with \\
%}%


\begin{abstract}
A vast array of phenomena, ranging from chemical reactions to phase transformations, are analysed in terms of a free energy surface defined with respect to a single or multiple order parameters. Enhanced sampling methods are typically used, especially in the presence of large free energy barriers, to estimate free energies using biasing protocols and sampling of transition paths. Kinetic reconstructions of free energy barriers of intermediate height have been performed, with respect to a single order parameter, employing the steady state properties of unconstrained simulation trajectories when barrier crossing is achievable with reasonable computational effort. Considering such cases, we describe a method to estimate free energy surfaces with respect to multiple order parameters from a steady state ensemble of trajectories. The approach applies to cases where the transition rates between pairs of order parameter values considered is not affected by the presence of an absorbing boundary, whereas the macroscopic fluxes and sampling probabilities are. We demonstrate the applicability of our prescription on different test cases of random walkers executing Brownian motion in order parameter space with an underlying (free) energy landscape and discuss strategies to improve numerical estimates of the fluxes and sampling. We next use this approach to  reconstruct the free energy surface for supercooled liquid silicon with respect to the degree of crystallinity and density, from unconstrained molecular dynamics simulations, and obtain results quantitatively consistent with earlier results from umbrella sampling.
% \YGcom{Un-commented without changes}
% }
\end{abstract}

\maketitle

% \begin{quotation}
% The ``lead paragraph'' is encapsulated with the \LaTeX\ 
% \verb+quotation+ environment and is formatted as a single paragraph before the first section heading. 
% (The \verb+quotation+ environment reverts to its usual meaning after the first sectioning command.) 
% Note that numbered references are allowed in the lead paragraph.
% %
% The lead paragraph will only be found in an article being prepared for the journal \textit{Chaos}.
% \end{quotation}

\section{Introduction}
Numerical free energy calculations have contributed immensely to our understanding of phase transitions and activated processes. In particular, determining the underlying landscape allows one to quantify the relative stability of the various states a system can exist in and also to probabilistically predict the time evolution of the system.
A specific area of interest to the present work in which numerical free energy calculations have contributed immensely to our understanding is that of polyamorphism in liquids\cite{StanleyPolymorphism}. 
Network-forming liquids such as water\cite{palmer2014metastable,debenedetti2020second}, silica\cite{chen2017liquid}, silicon\cite{vasisht2013liquid,goswami2022liquid}  and other model liquids liquids\cite{smallenburg2014erasing,ricci2017free} have been shown to exhibit multiple metastable liquid states, with an associated liquid-liquid phase transition, that are present alongside the globally stable crystalline state.
The deeply supercooled conditions at which these systems exhibit this liquid-liquid phase transition pose a challenge not just in experimental investigations but also in performing simulations. At these conditions, crystallisation can occur very rapidly (as in the case of silicon, which we address in this work), as a result of which the sampling of the metastable liquid state is very poor and enhanced sampling techniques such as umbrella sampling\cite{torrie1977nonphysical} need to be employed to reconstruct the free energy landscape.

A large number of the cases in which such free energy calculations are performed belong to the class of ``rare events", where a direct observation of the event is often unfeasible, even numerically. In this context, many free energy calculation methods have been devised to obtain accurate estimates of the high free energy barrier and to sample the transition state at the top of the barrier\cite{bennett1977molecular,torrie1977nonphysical,van2003novel,allen2006simulating,valsson2016enhancing}. These enhanced sampling techniques are used to determine the free energy landscape by improving the sampling efficiency in poorly visited regions of order parameter space and imposing a condition of equilibrium or zero net flux in the order parameter space.
Key to such numerical investigations is a suitable model of the system of interest as well as a low-dimensional representation with appropriately chosen collective variables, or order parameters, that effectively distinguishes the relevant states\cite{peters2013reaction}. The landscape as well as the probabilistic rate of the transformation are closely related, with the set of activated processes -- where free energy barriers separate the states of interest -- receiving tremendous scientific interest over many years. To a first approximation, the rate and the barrier height can be related through an Arrhenius-like equation in the following way:
\begin{equation}
k=Ae^{-\frac{\Delta G}{k_BT}}.
\end{equation}
Here, $k$ is the rate, $A$ is a kinetic pre-factor, $k_B$ is Boltzmann's constant, $T$ the temperature and $\Delta G$ the height of the barrier measured from the initial or reactant state. 
A prominent concept in this context is the mean first passage time, which is the inverse of the rate.  Kramers framed the progress of a reaction as a diffusive barrier crossing where the time evolution of the collective variable of interest obeyed the Smoluchowski equation \cite{kramers1940brownian}. A key assumption in this approach is that degrees of freedom other than the collective variable relaxes on timescales much shorter than the barrier crossing timescale, and can thus be averaged out. The connection between the mean first passage time and the free energy barrier can be directly exploited to provide estimates of the free energy from unconstrained simulations where the collective variable evolves from an initial value between a reflecting and an absorbing boundary, through a kinetic reconstruction, developed by Reguera and co-workers \cite{wedekind2007new,wedekind2008kinetic}. Such an approach does not require one to produce a condition of zero net flux and has been used in the context of metastable-to-stable phase transitions such as nucleation in deeply supercooled conditions\cite{wedekind2009crossover,lundrigan2009test,thapar2015simultaneous,goswami2021thermodynamics}. 
However, the above approach, based on the exact relationship between the free energy function and the mean first passage times, is available when one considers only a single order parameter. 

A number of researchers have noted the connection between equilibrium free energies and rates and their non-equilibrium steady state counterparts in driven or dissipative systems conditions\cite{crooks1998nonequilibrium,banik2000generalized,hummer2001free}. 
Framing the problem of free energy calculations in the case where a constant flux to an absorbing state alters the steady state sampling probability is relevant to the context of the liquid-liquid phase transition, which has been vigorously investigated since it was first proposed as an explanation for the thermodynamic anomalies exhibited by water based on numerical evidence\cite{poole1992phase}.
Biased simulations such as umbrella sampling have been extensively used to perform such free energy calculations. However, the choice of order parameter and bias protocol are key to obtaining meaningful results, with inappropriate choices leading to qualitatively misleading results\cite{goswami2021thermodynamics}.
Thus, a method to compute free energies from unconstrained simulations initialised from the disordered liquid and proceeding till crystallisation occurs would be of value both in the context of metastable network-forming liquids, as well as numerous other contexts where secondary or tertiary order parameters relax on comparable timescales to the primary order parameter separating metastable states from the globally stable state\cite{defever2019contour,verma2022proton}. 

In this work, we develop a methodology to reconstruct multi-dimensional free energy landscapes from unconstrained simulations evolving between reflecting and absorbing boundary conditions separated by a free energy barrier along a primary collective variable. We focus on reconstruction in cases where multiple metastable states exist, separated by a barrier along an orthogonal collective variable. We begin by defining the phenomenological rate of transformation from reactant, $A$, to product, $B$, as the ratio of two partition functions\cite{chandler1978statistical,bolhuis2002transition} weighted on paths connecting $A$ and $B$ and all paths exiting $A$, respectively. We consider the effect of including an additional absorbing condition and discuss conditions under which the rate is not altered. However, the effective positive flux\cite{van2005elaborating} between different regions of order parameter space and the steady state sampling, which are respectively related to the two aforementioned partition functions, are. Using this, we are able to relate the steady state sampling in the presence of the additional boundary (through which there is a finite flux) to the underlying equilibrium sampling in the flux-balanced condition in the absence of the additional absorbing boundary. This relationship between steady state sampling and equilibrium sampling enables an estimate of the free energy landscape from unconstrained trajectories. We demonstrate the efficacy of this approach on a model of independent overdamped Brownian random walkers on a potential energy surface as well as on the reconstruction of the free energy surface for supercooled Stillinger-Weber silicon\cite{stillinger1985computer} from unconstrained molecular dynamics trajectories. Results are compared with recently published estimates of the free energy landscape for silicon from umbrella sampling simulations\cite{goswami2021thermodynamics,goswami2022liquid}.

The paper is organised in the following way: in Section~\ref{sec:extending_2d} we discuss how one can extend the single order parameter free energy to multiple order parameters under the assumption of Boltzmann sampling along the other order parameters. In Section~\ref{sec:systems_studied} we describe the systems on which we employ our approach, the test system of independent random walkers on a potential energy landscape in Section~\ref{subsec:test_system}, as well as liquid silicon in Section ~\ref{subsec:silicon_procedure}. In Section~\ref{sec:MFPT} we describe briefly the mean first passage time (MFPT) method employed to reconstruct one dimensional free energy profiles. In Section~\ref{sec:ortho_OP_sample}, we describe the reconstruction of free energies as a function of multiple order parameters using the MFPT method and the assumption of Boltzmann sampling along the second order parameter, which reveals the inadequacies of such an approach. In Section~\ref{sec:rates_and_fluxes} we derive the relationship between steady state sampling and equilibrium sampling in a multi-dimensional order parameter space, which is our main result.  In Section~\ref{sec:toy_model_tests}, we describe results on the test system demonstrating the effectiveness of our approach. We then apply this method to the more complex case of supercooled liquid silicon in Section~\ref{sec:silicon_results} and reconstruct the barrier to crystal nucleation as well as the barrier profile along a second order parameter, density, revealing the presence of two liquid states. In Section~\ref{sec:improving_estimates} we discuss possible ways to improve on our approach, focusing on a few key shortcomings before a brief discussion in Section~\ref{sec:discussion} summarising our findings and promising future directions.

% \sscom{Check and correct section numbers} 
% \YGcom{fixed}.
% In using a single order parameter description with a diffusive barrier crossing, a number of caveats need to be considered which have been discussed in various works. Such considerations include anisotropic diffusivity of the order parameters, multiple routes for the transition and multiple metastable states. As discussed by Peters and others, the reaction prefers the minimum energy path, however, this is often treated as an assumption. Some works do investigate the role of secondary order parameters, such as Bagchi and cluster quality guys but the role of one order parameter is pre-eminent.

\section{Extending the one order parameter free energy to multiple order parameters}\label{sec:extending_2d}
Here, the steps to obtain the two order parameter free energy $\beta \Delta G(x,y)$ from the single order parameter free energy $\beta \Delta G(x)$ and the sampling along two order parameters, $P(x,y)$, are described.
In equilibrium the sampling probability can be related to free energy differences in the following way,
\begin{equation}
P_{eq}(x,y) = A e^{-\beta \Delta G(x,y)}
\end{equation}
Using the single order parameter sampling probability,
\begin{equation}
P(x) = \int_{-\infty}^{\infty}P(x,y)dy
\end{equation}
%{\color{blue}
we can write,
\begin{equation}
P_{eq}(x) = \int_{-\infty}^{\infty}P_{eq}(x,y)dy = A\int_{-\infty}^{\infty}e^{-\beta \Delta G(x,y)}dy = A e^{-\beta \Delta G(x)}
\end{equation}
Using this relation between the free energy along $x$, $\beta\Delta G(x)$ and $P_{eq}(x)$ to give
% The free energy along $x$, $\beta\Delta G(x)$, can thus be related to the quantity $P_{eq}(x)$ through the normalisation factor, which we can write as:
% \begin{equation}
% A = P_{eq}(x)e^{\beta \Delta G(x)}
% \end{equation}
% giving,
\begin{equation}
P_{eq}(x,y) = P_{eq}(x)e^{\beta \Delta G(x)}e^{-\beta \Delta G(x,y)}
\end{equation}
From this, one can rearrange and to get
% \begin{equation}
% e^{\beta \Delta G(x) - \beta\Delta G(x,y)}= \frac{P_{eq}(x,y)}{P_{eq}(x)}
% \end{equation}
% or, 
\begin{equation}
\beta\Delta G(x,y) = \beta \Delta G(x) - ln\left( \frac{P_{eq}(x,y)}{P_{eq}(x)}\right)
\label{eq:bdgxy}
\end{equation}
% \begin{align}
% \beta \Delta G(x,y) - \beta \Delta G(x) &= - ln\left( \frac{P_{eq}(x,y)}{P_{eq}(x)}\right) \nonumber \\
% \beta\Delta G(x,y) 
% &= \beta \Delta G(x) - ln\left( \frac{P_{eq}(x,y)}{P_{eq}(x)}\right)
% \end{align}
Eq.~\ref{eq:bdgxy} is a relation between free energy and equilibrium probabilities.
Note that if the relative weights of sampling different $y$ for a given $x$, $P_{st}(y;x)$, are in equilibrium, then we can substitute $P_{eq}(x,y)$ with the measured $P_{st}(y;x)$ in Eq.~\ref{eq:bdgxy} to obtain $\beta\Delta G(x,y)$. 
%However, we note that so long as the conditional probability, $P_{st}(y;x)$, of sampling different $y$ for a given $x$ is equilibrium, 
We need to define $P_{st}(x) = \int dy P_{st}(y;x)$ for the denominator in Eq.~\ref{eq:bdgxy} to do so. In this case $\beta\Delta G(x)$ needs to be obtained independently, from some other method like umbrella sampling along $x$ or the kinetic reconstruction from the mean first passage time\cite{wedekind2007new,wedekind2008kinetic,wedekind2009crossover}. 
%}
One can compare the measured free energy from (say) the single order parameter reconstruction along $x$, %Eq.~\ref{eq:bdgx}
, with the quantity $G(x)$ which is given by:
\begin{equation}
e^{- \beta \Delta G(x)} = \int_{-\infty}^{\infty}e^{-\beta \Delta  G(x,y)}dy
\label{eq:vpx}
\end{equation}
% \begin{align}
% B(x) &= -\frac{1}{P_{st}(x)}\left [ \int_{x}^{b} P_{st}(x')dx' - \frac{\tau(b)-\tau(x)}{\tau(b)}\right] \nonumber \\
% \beta \Delta G(x) &= \beta \Delta G(x=1) + ln \left ( \frac{B(x)}{B(1)}\right ) - \int_{1}^{x} \frac{dx'}{B(x')}
% \label{eq:bdgx}
% \end{align}
% One should therefore get $\Delta G(x) = G_p(x)$.
Eq.~\ref{eq:bdgxy} is the result that allows the extension of single order parameter free energies to multiple order parameters. However, as we will see, this can be used as-is only when sampling along the other order parameters is Boltzmann. When this is not the case, corrections need to be used to obtain the correct free energy, which are discussed in detail in Sec.~\ref{sec:rates_and_fluxes}. As we shall see in the discussion that follows, one can obtain the free energy as a function of one or multiple order parameters by employing the identified corrections.

\section{Systems studied}\label{sec:systems_studied}

In this section we describe the systems on which we test this method. We first test the numerical reconstruction procedure for the case of independent overdamped random walkers on an energy landscape which mimic the behaviour of trajectories in order parameter space for which the dynamics are not discontinuous. We consider first a landscape having two metastable minima and a globally stable minimum, motivated by the problem of reconstructing the free energy landscape of liquids displaying polyamorphism. We also consider other such test cases (see Appendix~\ref{appsec:alt_potentials}) as well as supercooled liquid silicon.

\subsection{Test system}\label{subsec:test_system}
The model potential energy function we consider is of the form in Eq.~\ref{eq:vxy}, which is a sum of $4$ Gaussian functions in two dimensions (see Fig~\ref{fig:vxy_form} for illustration and Table \ref{table:pot_details} in Appendix~\ref{appsec:potential} for values of constants). An additional harmonic cost potential, $V_c(y)$ is applied to ensure that random walkers sample order parameter space within $y \in [-0.5,0.5]$. 
%The potential energy function is equivalent to the free energy landscape explored by trajectories in order parameter space. 
\begin{equation}
V(x,y)=\sum\limits_{i=1}^{4}V_i(x,y;x_i,y_i,\sigma x_i,\sigma y_i) + V_c(y)
%+V_2(x,y) +V_3(x,y) +V_4(x,y)
\label{eq:vxy}
\end{equation}
\begin{figure}[h!]
\centering
\includegraphics[trim=30 0 5 30,clip,scale=0.6]{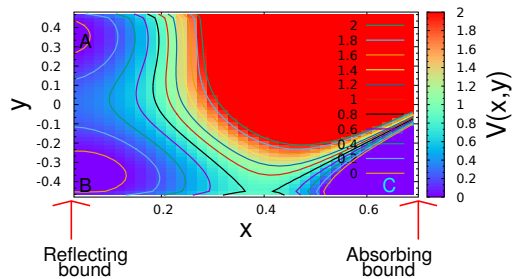}
\caption{The model potential energy landscape for Eq.~\ref{eq:vxy}. Shown here for a case where the barrier along $x$ is approximately $2~k_BT$. Contour lines are $0.25~k_BT$ apart. The reflecting boundary is at $x=0$ and absorbing boundary at $x=0.75$ as shown. In order to determine the accuracy of sampling along $y$, we compare slices along different values of $x$, such as $x~\in~[0.10,0.12]$ as marked in the figure.}
\label{fig:vxy_form}
\end{figure}
The surface is specified such that there are two saddles, with two metastable states separated from the globally stable state at large $x$ by the barrier along $x$ at $x\approx 0.4$.
%\sscom{$x\approx 0.4$}
Such a surface is relevant in contexts such as liquid polymorphism, where multiple metastable liquid states co-exist with the globally stable crystalline state in a number of anomalous model liquids\cite{palmer2014metastable,smallenburg2014erasing,debenedetti2020second,goswami2022liquid}.
The flux to the globally stable state at $x>0.4$ is controlled by the height of the scaled barrier (in units of $k_BT$) along $x$. We study cases where the height of the scaled barrier is low, $2~k_BT$ to $6~k_BT$, where the flux across the barrier along $x$ is high, leading to larger deviations of steady state sampling from equilibrium sampling.
% \sscom{The description before is unnecessarily complicated and confusing. You have a potential and you vary T. See if you can easily rephrase that way.} 
% \YGcom{We do also vary the height of the barrier directly by changing the barrier height at constant temperature, e.g. Fig 4, so maybe the new (current) version should be ok? }
This is also consistent with our expectation for deeply supercooled liquids where the barrier to crystallisation is found to be very low\cite{goswami2021thermodynamics}.
$N_{traj}$ non-interacting random walkers are initialised at $x=0$ and different $y$, either sampling the Boltzmann distribution or at a specified $y$ value at $x=0$.
A reflecting boundary condition is placed at $x=0$ and an absorbing boundary condition at $x=x_c=0.75$
% \YGcom{
% $x_c$ introduced here} \sscom{Let us call it $x_c$ to avoid confusion.}
for all $y$. For the reflecting boundary condition at $x=0$, if a trial move places a particle at $x^{'}_{new}<0$, the trial move is modified to $x_{new}=-x^{'}_{new}$ and accepted or rejected depending on the Boltzmann-weighted energy at $\beta\Delta V(x_{new},y_{new})$.
No boundary conditions are imposed along the $y-axis$. 
% \sscom{The following seems unnecessary and complicated. You would be considering each walker independently. Rephrase.}
% \YGcom{
Each MC sweep consists of $N_{traj}$ trial displacements of the random walkers. In each trial displacement, a random walker is chosen with uniform probability and is displaced by  ${dx,dy}\in[-\delta:+\delta,-\delta:+\delta]$
% }
% In each MC sweep, $N_{traj}$ walkers are chosen with uniform random probability and displaced by a .
Here, the value of $\delta=4\times10^{-3}$ is used while the order parameter space is divided into equal-sized square bins of size $0.02$. While using larger step sizes, i.e., comparable to the bin size, introduces sampling issues, we have determined that the chosen step size does not affect our results.
Trial displacements are accepted or rejected using a Boltzmann weight for the change in energy for every trial move. One can thus obtain the steady state sampling probability, $P_{st}(x,y)$, from a number of trajectories that proceed to an absorbing boundary condition. We demonstrate free energy reconstruction using $N_{traj}=600$ such independent trajectories. We also compare results for the reconstruction of the barrier along $x$ with those obtained for the same set of trajectories by a kinetic reconstruction using the MFPT developed by Reguera and co-workers\cite{wedekind2007new,wedekind2008kinetic}. The procedure for this method and results for the single order parameter problem are discussed in the next section, following which we discuss the two order parameter reconstruction using Eq.~\ref{eq:bdgxy}.
\subsection{Supercooled silicon}\label{subsec:silicon_procedure}
The other system we consider is liquid silicon modelled by the Stillinger-Weber potential\cite{stillinger1985computer}.
The existence of two metastable liquid states for this model has been investigated intensely\cite{sastry2003liquid,ganesh2009liquid,beye2010liquid,vasisht2011liquid} with recent free energy calculations also finding a high density liquid and a low density liquid separated by a free energy barrier\cite{goswami2022liquid}. This scenario is analogous to other network-forming liquids such as water\cite{palmer2014metastable,debenedetti2020second}, silica\cite{chen2017liquid} and patchy colloidal model liquids\cite{smallenburg2014erasing,neophytou2022topological} where two liquid states have been identified. In the case of water, silica and silicon, the globally stable crystalline state is separated from two metastable liquid states by the free energy barrier to crystallisation. In order to reconstruct the free energy from unconstrained molecular dynamics simulations, we initialise $400$ independent molecular dynamics simulations from configurations of randomly placed particles without overlap at a density of $2.48~gcc^{-1}$. Molecular dynamics simulations are performed in the isothermal-isobaric ensemble using the LAMMPS package\cite{plimpton1995fast} at target pressures and temperatures of $P=0.75~GPa$ and $T=975K,~985K,~995K$, monitoring the size of the largest crystalline cluster\cite{romano2011crystallization,goswami2021thermodynamics}, denoted $n_{max}$, and the density ($\rho$), with simulations being extended till a largest cluster size of $n_{max}=80$ being reached. Trajectories in ($n_{max}$,$\rho$) space are treated as random walks on the underlying free energy surface. 

\section{Kinetic reconstruction along one order parameter using the mean first passage time}\label{sec:MFPT}
We use Eq.~\ref{eq:bdgx} and Eq.~\ref{eq:Bx} as described in\cite{wedekind2007new,wedekind2008kinetic,wedekind2009crossover,wedekind2015optimization} to obtain the 1D barrier along $x$ from a set of unconstrained trajectories that proceed until the absorbing boundary at $x=0.75$ is reached.
\begin{equation}
\beta \Delta G(x) = \beta \Delta G(x=1) + ln \left ( \frac{B(x)}{B(1)}\right ) - \int_{1}^{x} \frac{dx'}{B(x')}
\label{eq:bdgx}
\end{equation}
\begin{equation}
B(x) = -\frac{1}{P_{st}(x)}\left [ \int_{x}^{b} P_{st}(x')dx' - \frac{\tau(b)-\tau(x)}{\tau(b)}\right]
\label{eq:Bx}
\end{equation}
Eq.~\ref{eq:bdgx} and Eq.~\ref{eq:Bx} are the equations used to reconstruct the free energy from the MFPT and the steady state probability.
\begin{figure}[hpbt!]
\centering
\includegraphics[scale=0.4]{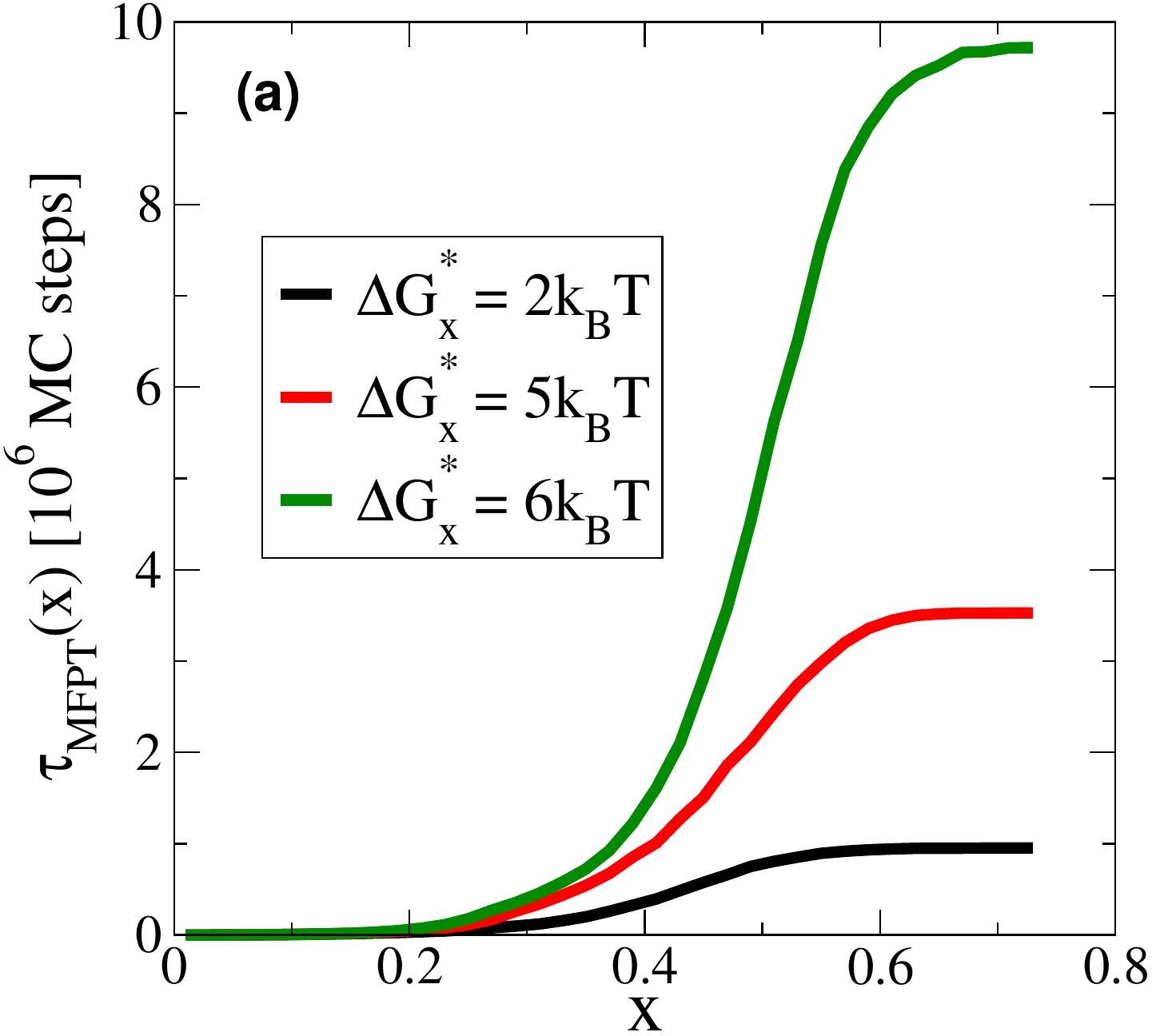}
\includegraphics[scale=0.4]{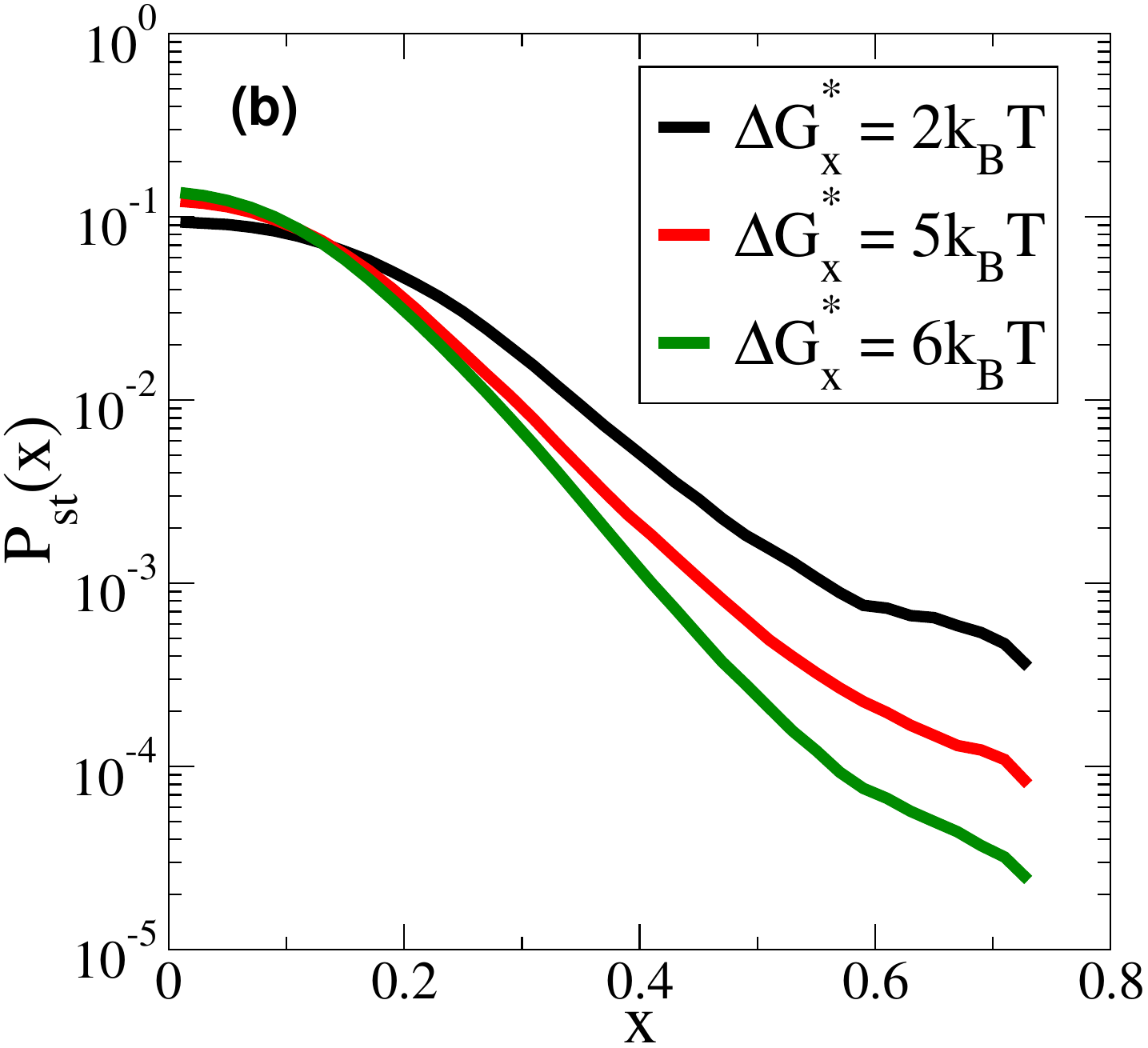}
\caption[Mean first passage time, $\tau_{MFPT}(x)$ and the steady state probability $P_{st}(x)$.]{Mean first passage time, $\tau_{MFPT}(x)$ and the steady state probability $P_{st}(x)$. Absorbing boundary condition at $x=0.75$, reflecting boundary condition at $x=0$. Number of walkers is $N_{traj}=600$, sufficient to generate smooth data.
% \sscom{Would be good to label dG/kT = xx. Here and elsewhere. But leave it if it is too much work.} 
% \YGcom{Have used this everywhere so would prefer not to change}
}
\label{fig:tau_Pvx}
\end{figure}

%\subsection{1D free energy barrier}
\begin{figure*}[htpb!]
\centering
\subfloat{\includegraphics[scale=0.4]{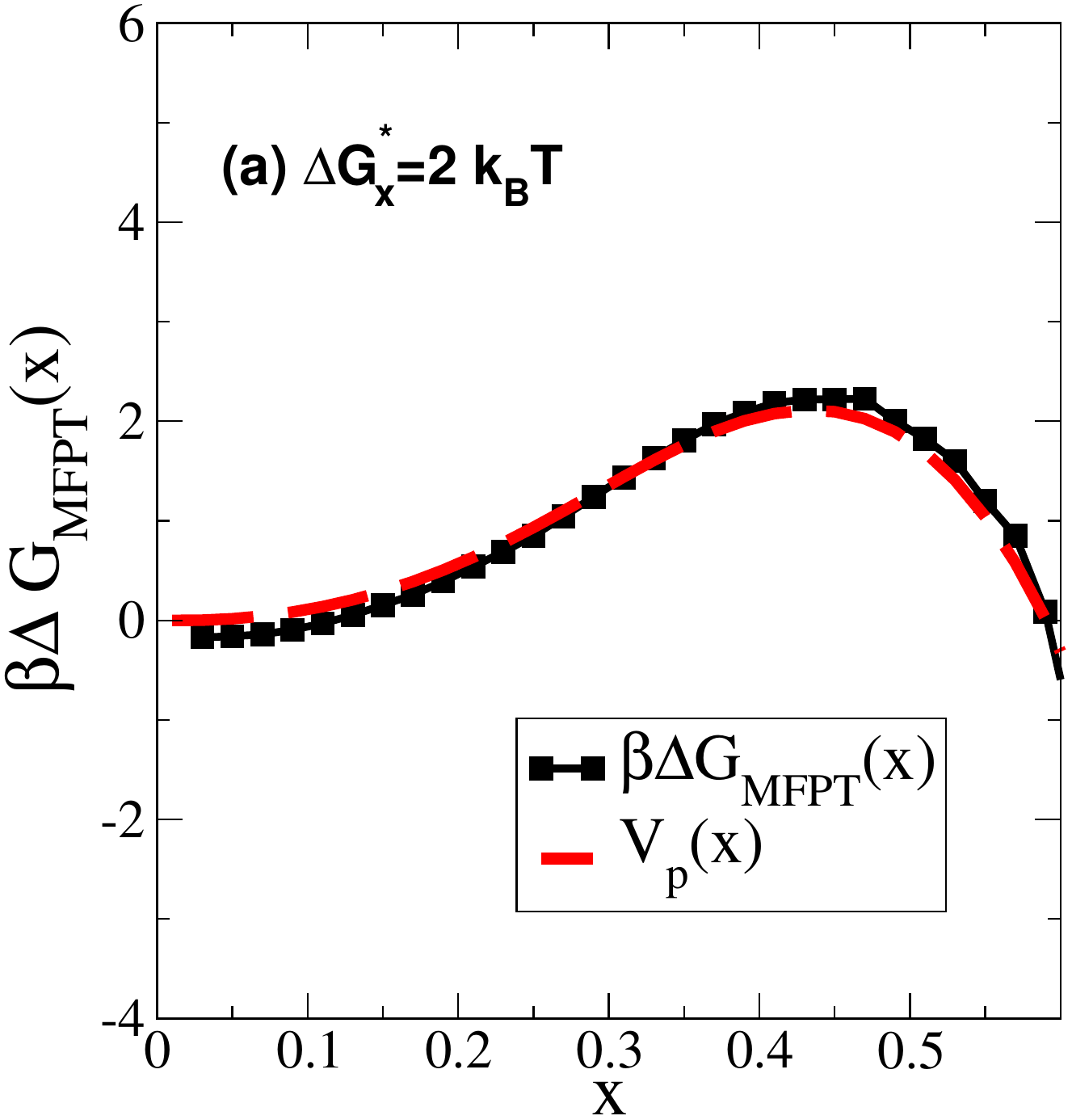}}
\subfloat{\includegraphics[scale=0.4]{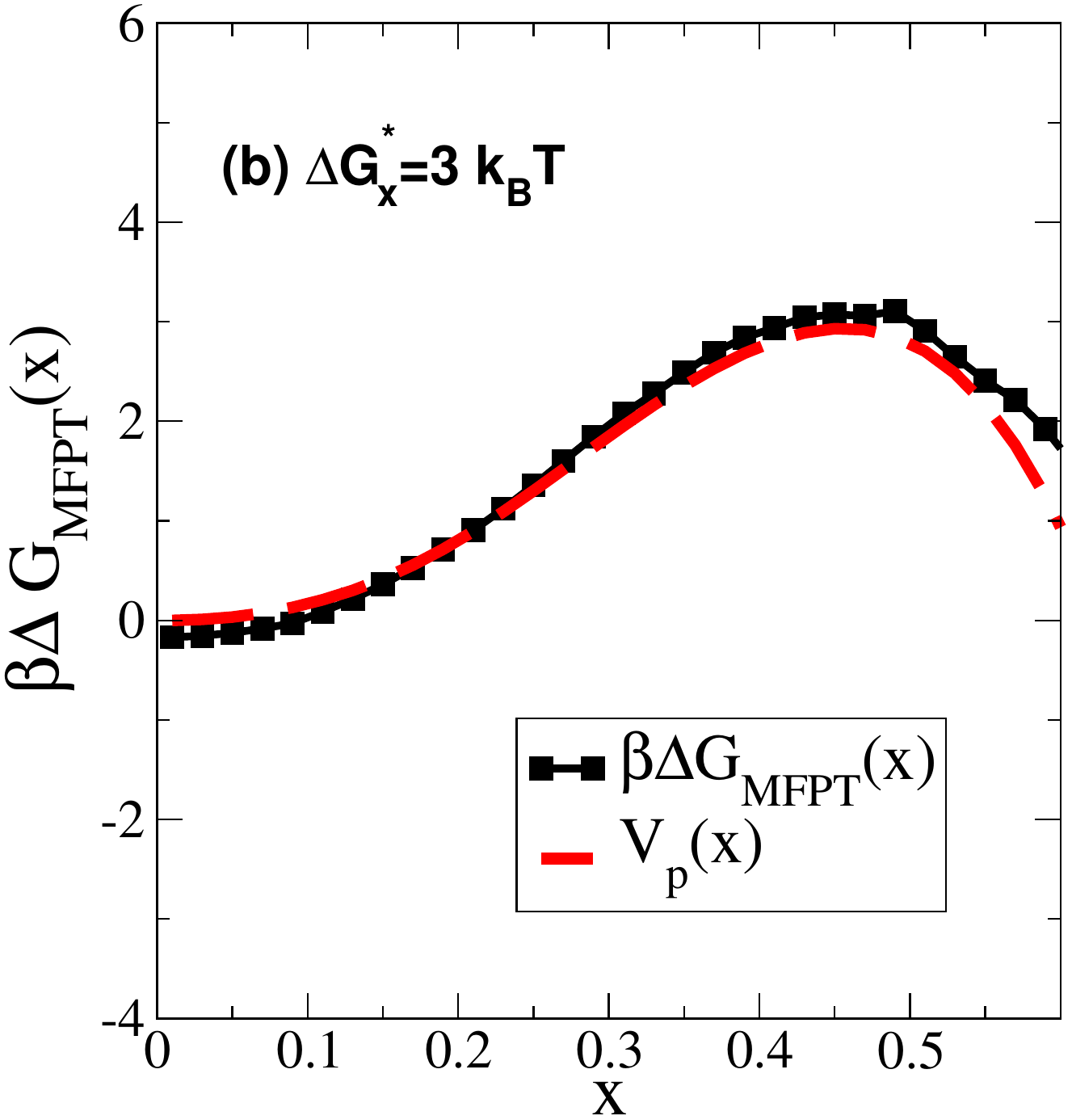}}
\subfloat{\includegraphics[scale=0.4]{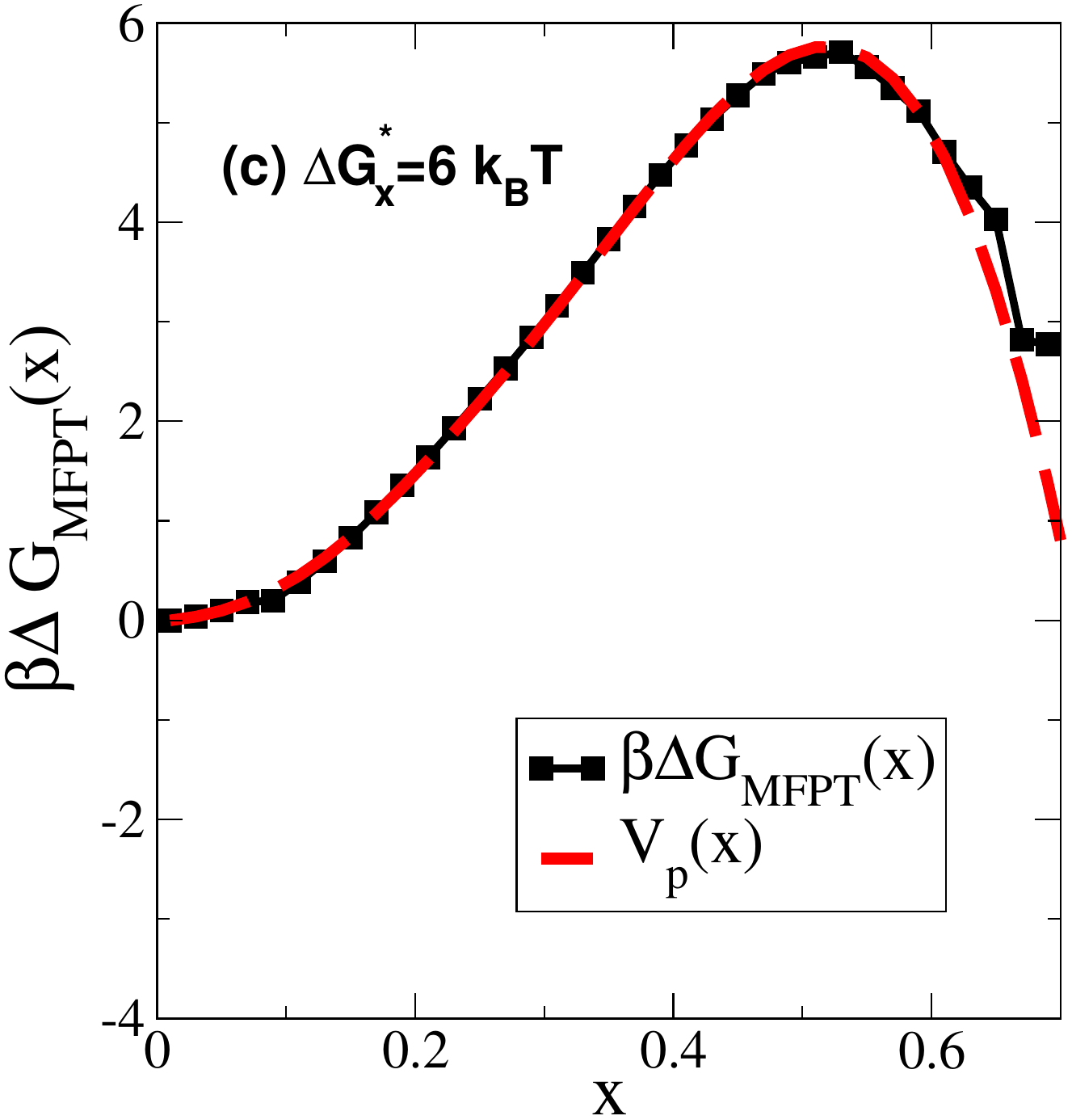}}
\caption{The free energy along $x$, $ \Delta G(x)$ obtained from Eq.~\ref{eq:bdgx}. For reference, the potential dependence is shown by integrating out the $y$ variation (Eq.~\ref{eq:vpx}). The number of walkers is $N_{traj}=600$ and the absorbing boundary condition is placed at $x=0.75$. Random walks are performed at a temperature of $T=0.08$ on $3$ landscapes with barrier heights along $x$ of $1-2~k_BT$ (panel ({\bf a})), $3~k_BT$ (panel ({\bf b})) and $5~k_BT$ (panel ({\bf c})). Error is minimised by shifting the curves to minimise the difference $|\beta \Delta G(x) - V_p(x)|$.
}
\label{fig:bDGx_mfpt}
\end{figure*}
Fig.~\ref{fig:tau_Pvx} show the measured mean first passage time and the steady state probability along $x$ for $3$ scaled barrier heights along $x$. The reconstructed free energy using Eq.~\ref{eq:bdgx} and Eq.~\ref{eq:Bx} is shown in Fig.~\ref{fig:bDGx_mfpt}, compared with the expected curve integrating out the $y$ dependence, given in Eq.~\ref{eq:vpx}. 

\section{Sampling along orthogonal order parameters: Deviation for high flux through the absorbing boundary}\label{sec:ortho_OP_sample}

\begin{figure}[htpb!]
\centering
\includegraphics[scale=0.45]{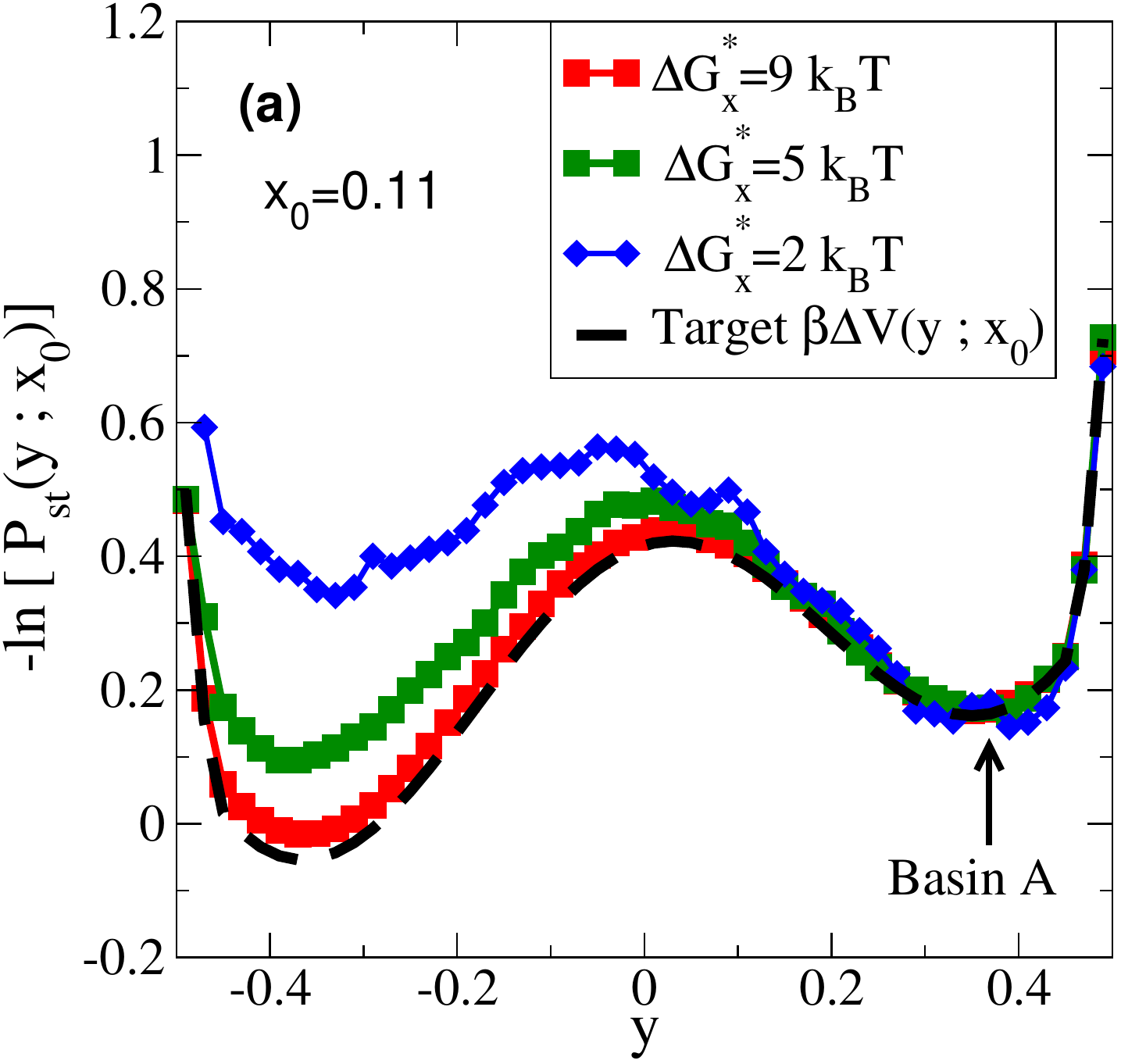}
\includegraphics[scale=0.45]{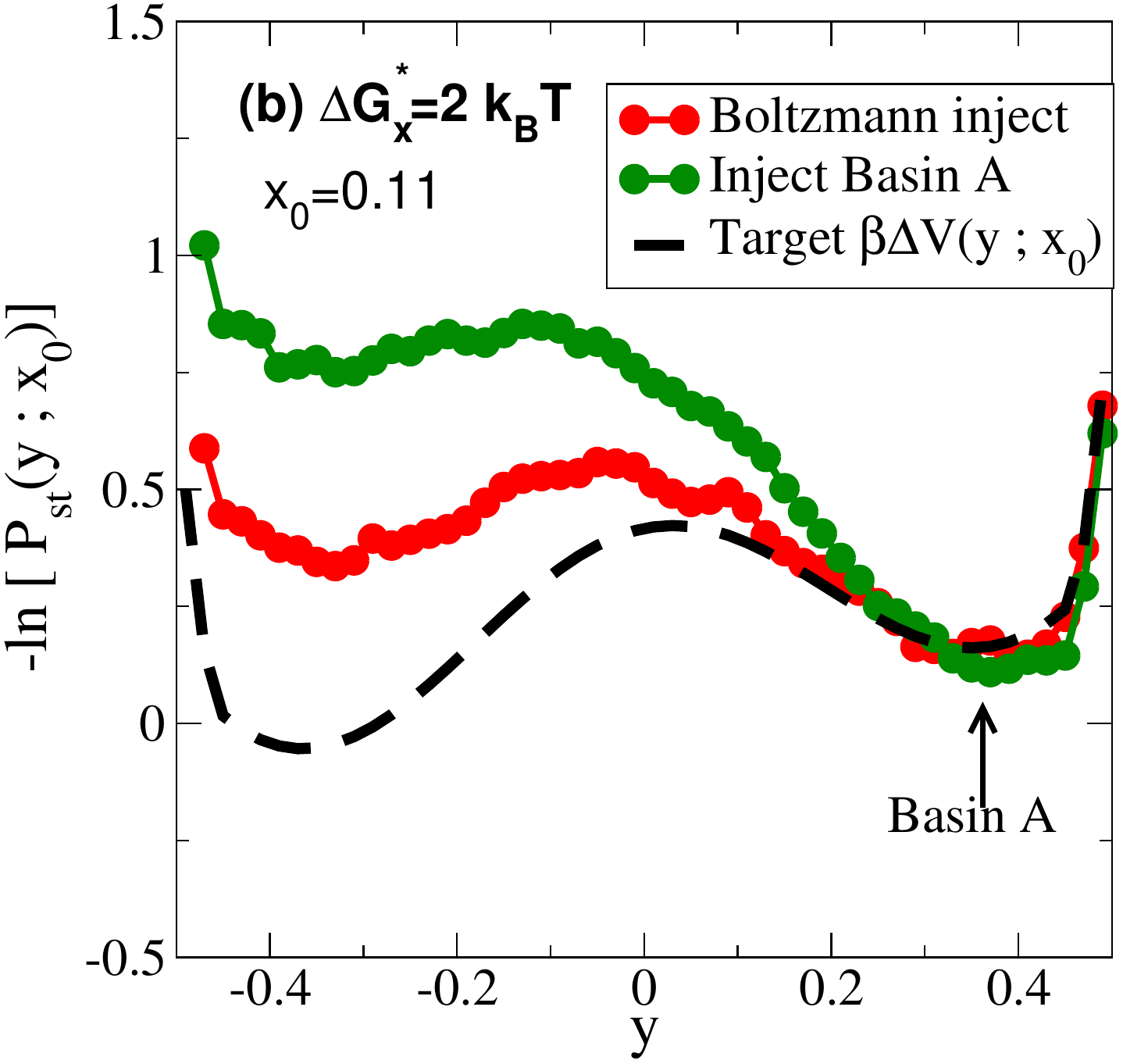}
\caption{Slices along $y$ of the negative log of the steady state probability, $-ln [P_{st}(y;x_0)]$, compared with the target, underlying landscape, $V(y;x_0)$ at $x_0~\in [0.10,0.12]$. $N_{traj}=600$ 
% \sscom{500 or 600? see previous figure caption.}
random walkers are initialised at $x=0$ and at different $y$ proportional to the Boltzmann weight on the target distribution (panel ({\bf a})) or at a point in the basin $A$ (panel ({\bf b})). As the scaled barrier along $x$ is lowered, the deviation in steady state sampling of $y$ from the target distribution increases as shown in panel ({\bf a}). In panel ({\bf b}) we observe that for a fixed value of the scaled barrier along $x$, the deviation of the steady state distribution from the target distribution is higher when the initialisation is at a single point along $y$. }
\label{fig:Py_slices_v_C}
\end{figure}

We next compare the reconstructed free energy along the $y$ direction to the corresponding cross-section of the potential along $y$ using Eq.~\ref{eq:bdgxy}. The height of the barrier is controlled by modulating the potential (see Appendix~\ref{appsec:potential}). We find that as the barrier along $x$ is lowered, the deviation of steady state sampling from the target, $V(y)_{x=0.01}$, increases. This is shown in Fig.~\ref{fig:Py_slices_v_C}. This can be rationalised as the enhanced flux across the lower barriers driving the system away from equilibrium sampling to a non-equilibrium steady state. 

The errors in the sampling along the orthogonal order parameter, $y$, that are summarised in Fig.~\ref{fig:Py_slices_v_C} arise from compounding factors:
Firstly, when the scaled barrier along $x$ is low, the net flux to the absorbing boundary prevents Boltzmann sampling along $y$ for $x<x^*$ (where $x^*$ is location of the barrier). The final steady state distribution also has a dependence on the injection probability, as seen in panel (b) of Fig.~\ref{fig:Py_slices_v_C} where the deviation is higher for the point injection compared to the Boltzmann weighted injection along $y$. We next discuss how to correct for these errors by identifying a relationship between steady state sampling and equilibrium sampling.

\section{Relating steady state sampling to equilibrium sampling}\label{sec:rates_and_fluxes}

We found in the previous section that sampling in the presence of a low barrier along $x$ and a constant non-zero flux across it, that the measured sampling along $y$ deviates from the underlying Boltzmann distribution.
This steady state is achieved by re-injecting or restarting a trajectory from an injection point between the reflecting and the absorbing condition and tracking it until it crosses the absorbing boundary, whereupon another trajectory is started, thus conserving the number of ``active" trajectories at any point of time.
% \sout{In this way we do not encounter effects of long time depletion.}
In this section we will discuss our approach to correct for the systematic deviation in sampling by considering the effect of the steady state flux across the absorbing boundary that is established.
In order to understand this, we begin by considering the rate(s) of traversal between any two regions of the order parameter space, $A$ and $B$, along the lines of the development of transition path sampling, transition interface sampling and forward flux sampling methods\cite{chandler1978statistical,dellago1998transition,bolhuis2002transition,van2005elaborating,allen2009forward,vanden2010transition}.
% \sscom{We represent the order parameter values for simplicity as $x$, and those of $A$ and $B$ by $x_A$ and $x_B$. We consider $A$ and $B$ to be points within the grid resolution we specify, but $x_A$ and $x_B$ can equivalently be considered to be a set of $x$ and $y$ values.} 
We represent the order parameter values for simplicity as $x$, and those of $A$ and $B$ by $x_A$ and $x_B$. We consider $A$ and $B$ to be points within the grid resolution we specify, but $x_A$ and $x_B$ can equivalently be considered to be a set of $x$ and $y$ values.
% \sout{consider order parameter points $A$ and $B$ is such that $x_A<x_B$, then}
We first write functions $h_A(x)$ and $h_B(x)$ where $h_A(x) = 1~if~x\in x_A$ and $0$ otherwise, and $h_B(x) = 1~if~x\in x_B$ and $0$ otherwise.
The phenomenological rate of transition from non-intersecting regions of order parameter space, reactant $A$ and product $B$, is then given in terms of the time correlation of the product of these functions as\cite{chandler1978statistical,frenkel2001understanding,bolhuis2002transition,van2005elaborating,allen2009forward}.
\begin{equation}
k_{AB} = \frac{d}{dt} C(t) = \frac{d}{dt} \left [ \frac{\langle h_A(x_0)h_B(x_t) \rangle}{\langle h_A(x_0) \rangle} \right ]
\label{eq:rate_from_correlator}
\end{equation}
The assumption of a time-invariant rate, related to the inverse mean first passage time, is typically invoked in the context of regions $A$ and $B$ that are separated by a barrier with a steady state rate across it that is established after an initial transient and decays on a global reaction timescale\cite{van1992stochastic,frenkel2001understanding}.
%\cite{kramers1940brownian,chandler1978statistical,muller1997rates,bolhuis2002transition,evans2020stochastic}.
Here, we consider a steady state scenario where the rate of traversal between any two non-overlapping regions of order parameter space is of interest.
The average here is over an ensemble of trajectories
% \sout{defined earlier,}
and weighted on the probability of observing a path connecting $A$ and $B$, of length $t$
% \YGcom{
, denoted $P[\{x_t\}]$. This is the probability of observing a trajectory, i.e., the sequence $\{x_0,x_1,\dots,x_t\}$
% }
For stochastic trajectories, with transition matrix $\mathbf{T}$,
\begin{equation}
P[\{x_t \}] = P(x_0)\prod\limits_{0<t'\leq t}\mathbf{T}_{x_{t'-1}x_{t'}}
\end{equation}
For the deterministic case the initial conditions fully specify the probability of observing a path, $P[\{x_t \}] = P(x_0)$.
In the equilibrium case, $\langle h_A(x) \rangle$ is equal to the equilibrium probability of being in $A$. 
Upon the introduction of an additional absorbing boundary $C$, one expects both the sampling probabilities as well as the probability of observing a given trajectory are altered. The probability of observing a path is altered through an alteration of the microscopic transition matrix $\mathbf{T}$, at some point $r$ in the vicinity of the absorbing boundary $C$. 
We can write that outward transition probabilities for a state neighbouring the new absorbing state, labelled $r$, are altered by the introduction of the new absorbing state $C$. For simplicity, we consider a single such state $r$; this choice should not affect our conclusions. The probability or weight of a path labelled $\{x_t\}$ becomes:
\begin{align}
P'[\{x_t \}] = &\prod\limits_{1<t'\leq t}\left [\mathbf{T}_{x_{t'-1}x_{t'}} + \delta_{r,t'-1}  \nonumber ( \mathbf{T'}_{x_{t'-1}x_{t'}} - \mathbf{T}_{x_{t'-1}x_{t'}}) \right ] \nonumber \\
&\times f(x_0)P(x_0)
\end{align}
Intuitively, the integral over all paths is changed when the fraction of paths between $A$ and $B$ that pass through $r$ is significant. Otherwise the term in the product remains unchanged. For this to be true, $A$ and $B$ should both be far from $C$, such that typical paths connecting them are significantly shorter than the typical length of a full trajectory that proceeds until it encounters $C$. 
Moreover, if the region $A$ is defined such that $f(x_0)$ is a constant value $f$, within $A$, then it can be ignored while evaluating the two integrals in Eq.~\ref{eq:rate_from_correlator}. Under these two conditions, we then assume that the phenomenological rates of traversal between $A$ and $B$, $k_{AB}$ and $k_{BA}$, are unaltered upon the addition of the absorbing state $C$. However, the flux between the two, as well as the steady state sampling probabilities are altered with respect to the corresponding equilibrium states. 
% \YGcom{
The flux from $A$ to $B$, $\langle \Phi_{AB} \rangle$, is defined here as the number of trajectories entering $B$ in a given time window $[t,t+\Delta t]$ that had their origin in $A$ at $t=0$. The flux per unit time is obtained by dividing it by the length of the interval, $\Delta t$\cite{van2003novel}. 
In the steady state condition, we can exploit the following replacement, $h_A(x_0)h_B(x_t)=h_A(x_{-t})h_B(x_0)$, whose time derivative can then be evaluated at $t=0$. Further, the time dependence for $h_A$ can also be dropped, given the steady state condition, so long as the trajectory did not visit $B$ prior to $t=0$ (to ensure that a trajectory that leaves $A$ once only counts towards the flux entering $B$ once)\cite{van2003novel,van2005elaborating}. 
Thus, the steady state flux, $\langle \Phi_{AB} \rangle = \langle d/dt|_{t=0} h_A(x_{-t})  h_B(x_0)\rangle$
% \sscom{(have removed the extra dot.)}
can be evaluated as the number of trajectories entering $B$ during some interval $\Delta t$, who were last in $A$ before $B$, and can be aggregated over a given time interval.
% }
%\sscom{Is there an integration over time backwards that needs to be mentioned?}
% \YGcom{
In the equilibrium case, the fluxes are balanced, and any random walker (or trajectory) that visits either $A$ or $B$, visits the other as well. The introduction of an additional absorbing boundary at $C$ introduces the added condition that only random walkers leaving $B$ ($A$) that reach $A$ ($B$) before reaching $C$ contribute to the flux, termed the splitting probability\cite{muller1997rates,huang2021random} in the non-equilibrium steady state case. 
% }
% \YGcom{
Thus, by matching rate of injection at $A$ to the rate of first passage at $C$, one obtains a steady state characterised by the macroscopic flux through $C$. One can then measure the number of trajectories that make a transition from $A$ to $B$, or the reverse, before reaching $C$, in the time it takes for $N$ such trajectories to traverse from the injection point to $C$. This gives the flux subject to the steady state specified by our injection rate.
% }
$\langle h_A \rangle$ is the probability that a randomly chosen starting point for a trajectory is in $A$, which is also the steady state sampling probability for $A$.
Using this, we are now able to write the phenomenological rate as:
\begin{equation}
k_{AB} = \frac{\langle \Phi_{AB}\rangle}{\langle h_{A}\rangle}.
\label{eq:kAB_phi}
\end{equation}
We now discuss how Eq.~\ref{eq:kAB_phi} can be used to relate the steady state sampling probability to the equilibrium sampling probability.
In equilibrium, $\langle \Phi_{AB}\rangle_{eq} = \langle \Phi_{BA}\rangle_{eq}$ (zero current), $\langle h_A \rangle_{eq} = P_{eq}(A)$, giving the detailed balance condition
\begin{equation}
k_{AB}P_{eq}(A) = \langle \Phi_{AB}\rangle_{eq} = \langle \Phi_{BA}\rangle_{eq} = k_{BA}P_{eq}(B)
\end{equation}
Upon addition of the absorbing boundary, $C$, trajectories exiting $A$ (or $B$) can now be terminated at $C$. In the resulting steady state condition, the probability that a randomly chosen starting point is in $A$ is now $\langle h_A \rangle_{st} \equiv P_{st}(A)$, altered from $\langle h_A \rangle_{eq}$.
In steady state, the fluxes $\langle \Phi_{AB}\rangle_{st}$ and $\langle \Phi_{BA}\rangle_{st}$ are not equal. In order to relate the steady state quantities to the equilibrium quantities, we first assume trajectories cannot be initiated at $C$ (the new absorbing state). The ratio of flux per unit time to sampling probability gives us the (assumed) unaltered rate.
Thus, in steady state, where $\langle \Phi_{AB}\rangle_{st} \neq \langle \Phi_{BA}\rangle_{st}$
\begin{align}
\frac{P_{st}(A)}{P_{st}(B)} &= \frac{\langle h_A \rangle_{st}}{\langle h_B \rangle_{st}} \nonumber \\
&= \frac{\langle \Phi_{AB}\rangle_{st}}{\langle \Phi_{BA}\rangle_{st}}\frac{k_{BA}}{k_{AB}} = \frac{\langle \Phi_{AB}\rangle_{st}}{\langle \Phi_{BA}\rangle_{st}}\frac{P_{eq}(A)}{P_{eq}(B)}
\end{align}
What we want is to infer the equilibrium sampling probability from the measured steady state sampling probability. It is helpful to then re-write the equation above as 
\begin{equation}
\frac{P_{eq}(B)}{P_{eq}(A)} = \frac{\langle \Phi_{AB}\rangle_{st}}{\langle \Phi_{BA}\rangle_{st}}\frac{P_{st}(B)}{P_{st}(A)}.
\label{eq:Pst_to_Peq_orig}
\end{equation}
The quantities on the right hand side are evaluated from numerical simulations, for all $B$ of interest, which them results in an estimate of the free energies relative to that at $A$. 

\subsection{Notes on numerical implementation}
We will drop the $\langle \rangle$ hereafter, when describing $\langle \Phi_{AB}\rangle$. 
We simulate $N_{traj}$ trajectories, injected at $x=0$ in the basin marked $A$ in Fig.~\ref{fig:vxy_form}, that proceed to the absorbing bound, marked $C$ in Fig.~\ref{fig:vxy_form}, through a random walk on the potential surface $V(x,y)$. 
We then obtain $P_{st}(x,y)$ from the cumulative number of times each trajectory visits each $(x,y)$ bin.
% \sscom{something missing here}
% Thus, for $N_{traj}$ trajectories of lengths $T_i$, and can be written as sequences $\{ x_i(t),y_i(t)\}$, we can write the sampling probability in the following way
% \begin{equation}
% P_{st}(x,y) = \frac{1}{\sum_i {T_i}}\sum\limits_{i=1}^{N_{traj}}\sum\limits_{t=1}^{T_i} \delta(x_i(t) - x)\delta (y_i(t)- y)
% \label{eq:Pst_calc}
% \end{equation}
% \YGcom{
The relevant normalisation factor is the cumulative length of the $N_{traj}$ trajectories.
% }
% \YGcom{
We also compute the fluxes to and from the injection point, $A$, and every other $(x,y)$ in order parameter space, denoted $B$ in Eq.~\ref{eq:Pst_to_Peq_orig}. 
It is important to note that the established steady state is subject to the injection point, $A$, and the definition of a trajectory, which is initialised at the injection point and is terminated at the absorbing boundary, $x_c$. The rate of injection is chosen to match the rate of termination in the cases we discuss.
%, maintaining the total current of trajectories to the absorbing boundary. 

%Preserving this definition of a trajectory, i.e., without the possibility of subdividing a trajectory into smaller segments that only traverse back-and-forth between $A$ and $B$ without reaching $C$, we then ensure that crossings from $A$ to $B$ (or the reverse) are only counted once per trajectory, at the first instance of crossing from $A$ to some $B$ or from a given $B$ to $A$ at a subsequent time. 
We count the number of trajectories that, having visited $A$ at some time $t$, subsequently visit $B$ (a given $(x,y)$ bin)  at some later time $t'>t$. 
% We divide this by $N_{traj}$ to define $\Phi_{A\rightarrow B}$. 
Likewise, we count each trajectory that, having visited a given $B$ at some time $t$, subsequently visits $A$ at a time $t'>t$ before being terminated at the absorbing boundary. 
% Note that for any $(x,y)$ in order parameter space, i.e., any $B$, that is visited by a trajectory for the first time at some time $t$, $\Phi_{B \rightarrow A}$ is updated at the next subsequent time at which that trajectory visits $A$ (if any). 
% }
% \sscom{Add note to explain why this is the correct procedure.} 
We consider this count as providing estimates of  $\Phi_{A \rightarrow B}$ and  $\Phi_{B \rightarrow A}$, upon division by the total trajectory lenght, which we need not explicitly consider since we are only  interested in the ratio  $\Phi_{A \rightarrow B}/\Phi_{B \rightarrow A}$.  
%Through this procedure, we obtain the steady state sampling for each $(x,y)$ bin as well as well as the fluxes to and from $A$ required to obtain the equilibrium sampling probability.
Note that $\Phi_{B\rightarrow A}$ is sampled poorly beyond the barrier and far enough beyond the barrier, this quantity goes to $0$.
% $so$ is specified as the point of injection (in basin A).
% \sscom{Some notational confusion -- A vs so, si vs $x_c$ etc -- can be avoided.}
% \YGcom{Using uniform notation of $A$, $B$ and $x_c$ (for sink), $N_{traj}$ for number of random walkers (equivalently, trajectories). Also added note that $B$ refers to any given $(x,y)$ region.}
% \sscom{previous needs more clarity.} 
% \YGcom{Clarified and moved to appropriate location at the end of this sub-section.}
The ratio $\hat\pi(x,y)=\frac{\Phi_{A\rightarrow xy}}{\Phi_{xy\rightarrow A}}$ is what we apply as a correction factor, 
% \sscom{
in order to obtain the free energies, using Eq. \ref{eq:Pst_to_Peq_orig}.
% }
% \sout{in the metastable basin, pre-barrier.}
\begin{equation}
\frac{P_{eq}(x,y)}{P_{eq}(A)} = \hat\pi(x,y)\frac{P_{st}(x,y)}{P_{st}(A)}
\label{eq:ratio_xy_so}
\end{equation}
Inferring $P_{eq}(x,y)$ from $P_{st}(x,y)$ as shown earlier allows us to use Eq.~\ref{eq:bdgxy} which we initially arrived at as the extension of the single order parameter free energy estimate to multiple order parameters. This is done in the following way:
\begin{align}
\beta\Delta G(x,y) =& \beta\Delta G(x)-ln\left ( \frac{P_{eq}(x,y)}{P_{eq}(x)}\right ) \nonumber \\
=& \beta\Delta G(x) -ln\left ( \frac{P_{eq}(x,y)P_{eq}(A)}{P_{eq}(A)P_{eq}(x)}\right ) \nonumber \\
=& \beta\Delta G(x) -ln\left ( \frac{\hat\pi(x,y)P_{st}(x,y)P_{eq}(A)}{P_{st}(A)P_{eq}(x)}\right ) \nonumber \\
\beta\Delta G(x,y)=& \beta\Delta G(x) -ln\left (\frac{\hat\pi(x,y)P_{st}(x,y)}{P_{eq}(x)} \right ) + const.
\label{eq:bdgxy_hatpi}
\end{align}
Above we have written terms dependent only on $A$ as an irrelevant constant. $P_{eq}(x)$ is defined as
\begin{align}
P_{eq}(x) &= \int\limits_{-\infty}^{\infty} P_{eq}(x,y) dy \nonumber \\
&= \frac{P_{eq}(A)}{P_{st}(A)}\int\limits_{-\infty}^{\infty} \hat\pi(x,y) P_{st}(x,y) dy,
\end{align}
using Eq.~\ref{eq:ratio_xy_so}. $P_{eq}(x)$ is thus obtained upto a multiplicative constant integrating out the $y$-dependence of $\hat\pi(x,y)P_{st}(x,y)$. We obtain $\beta\Delta G(x)$ independently 
% \sscom{
and use Eq.~\ref{eq:ratio_xy_so} to obtain the multi-dimensional free energy $\Delta G(x,y)$.
% }
% \sout{and identify $y$-dependent corrections as above.}

Other schemes can be developed to define pairs for which we can apply Eq.~\ref{eq:bdgxy_hatpi}. Results shown in the subsequent sections consider the definition of $A$ used above, as the point of injection of the trajectories.
% \YGcom{
For the case where trajectories are injected at $x=0$ with a Boltzmann-weighted injection probability at different $y$, the same point is chosen as the state $A$ as in the point-injection case (the basin $A$ in Fig.~\ref{fig:vxy_form}).
% }
\subsection{Similar results in the literature}
% The reconstruction of multi-dimensional free energy surfaces has received considerable attention recently.
The treatment of the phenomenological rate employed here has established itself as an immensely useful concept in numerical rate calculations and also free energy calculations in concert with milestoning, path and interface sampling\cite{bolhuis2002transition,van2005elaborating,allen2009forward}. 
% \sscom{refs}
An analysis of trajectory segments in partial-path transition interface sampling (PPTIS) can be related to the single order parameter equivalent of the result here\cite{qin2019efficient}. In this method also, backward fluxes beyond the barrier are not easily obtained.
Typical interface sampling and methods that enhance fluxes rely on creating a situation of equal forward and backward flux to achieve equilibrium sampling\cite{valeriani2007computing,qin2019efficient}. In this work, we find that one can use a finite, but unequal, backward flux to infer equilibrium sampling probability from the measured steady state sampling probability. Such a situation arises naturally where simulations proceed to an absorbing boundary. Forward flux sampling has also been combined with the mean first passage time, eliminating the need for backward trajectories and zero net flux\cite{thapar2015simultaneous}. 
Systematic alterations in the phenomenological rate have been investigated using a treatment of the rate expression that considers the effect of an additional field on the path partition functions, rather than altered boundary conditions\cite{kuznets2021dissipation}. 
Studies of first passage times and first passage probabilities for Markov processes with specific boundary conditions have discussed related concepts such as the splitting probability\cite{muller1997rates}. Of particular note is the study of random walks with stochastic resetting, where the effect of resetting on the mean first passage time and the rates has been investigated\cite{evans2020stochastic,huang2021random,chen2022first}. 
Generalisations of Kramers' formalism to open or driven systems in one dimension have also been described, with similar ideas\cite{banik2000generalized,hummer2001free}.

\section{Results for the test system}\label{sec:toy_model_tests}
\begin{figure*}[htpb!]
\centering
\includegraphics[height=5cm,width=5cm]{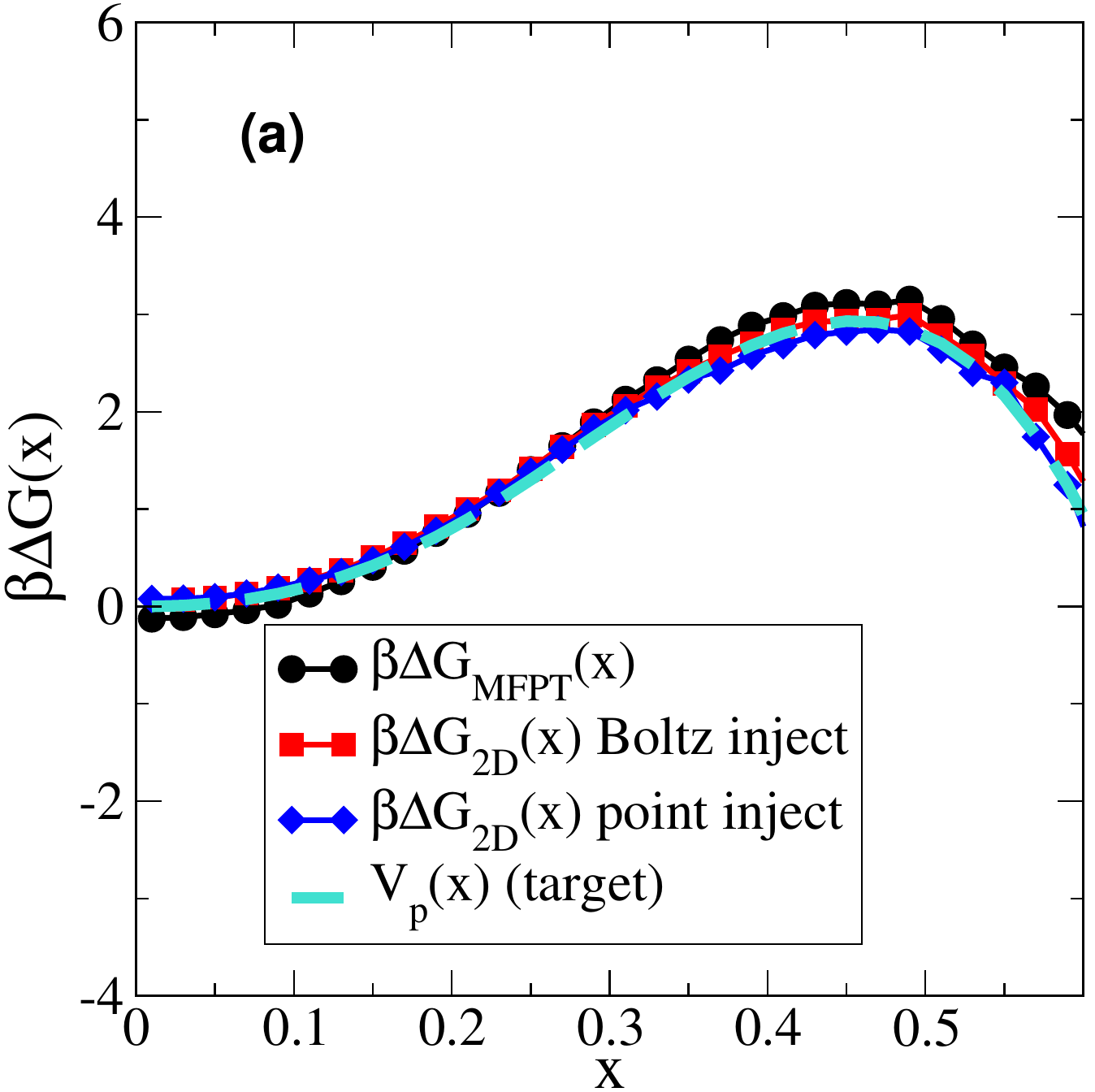}
\includegraphics[height=5cm,width=5cm]{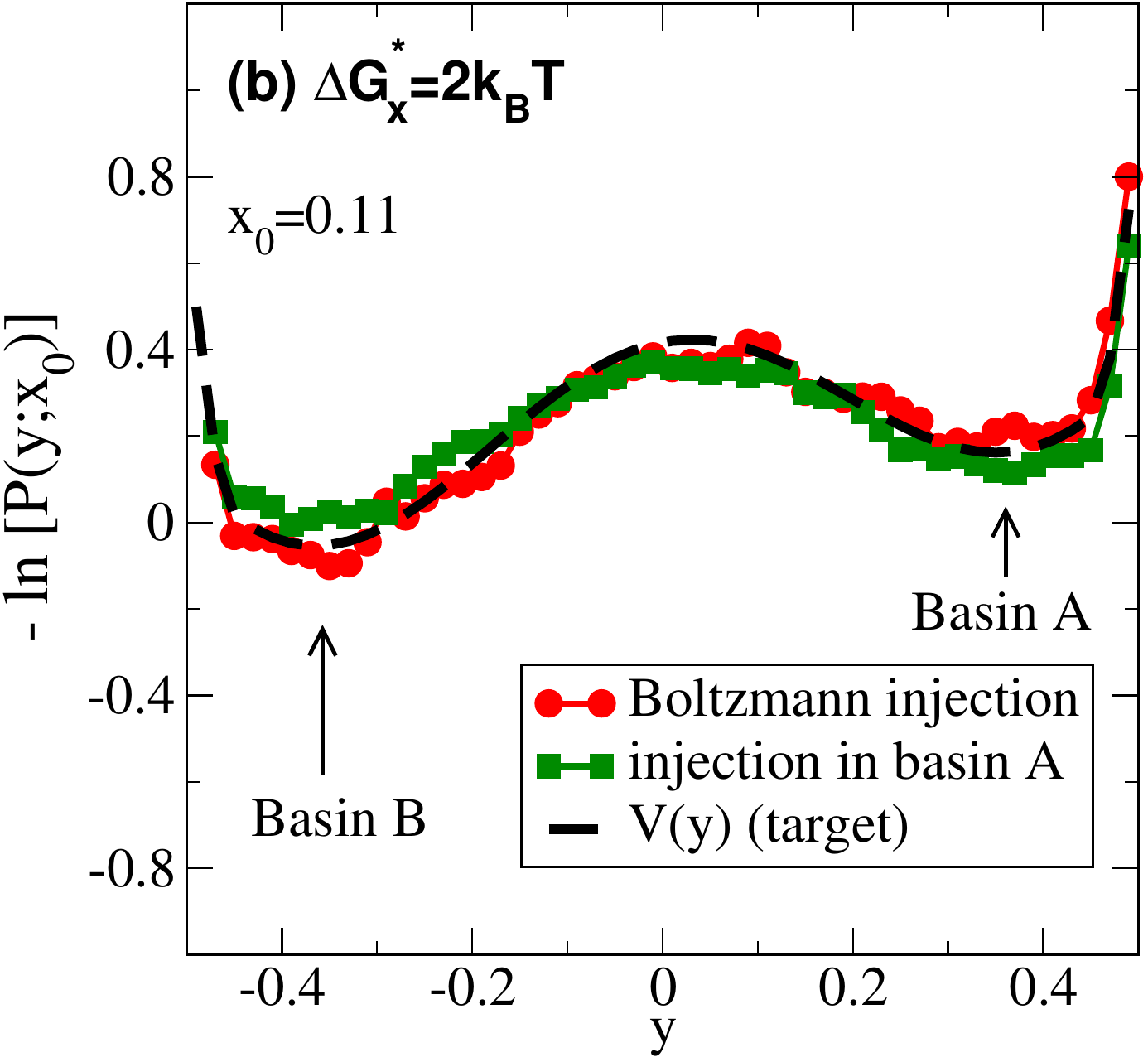}
\includegraphics[height=5cm,width=5cm]{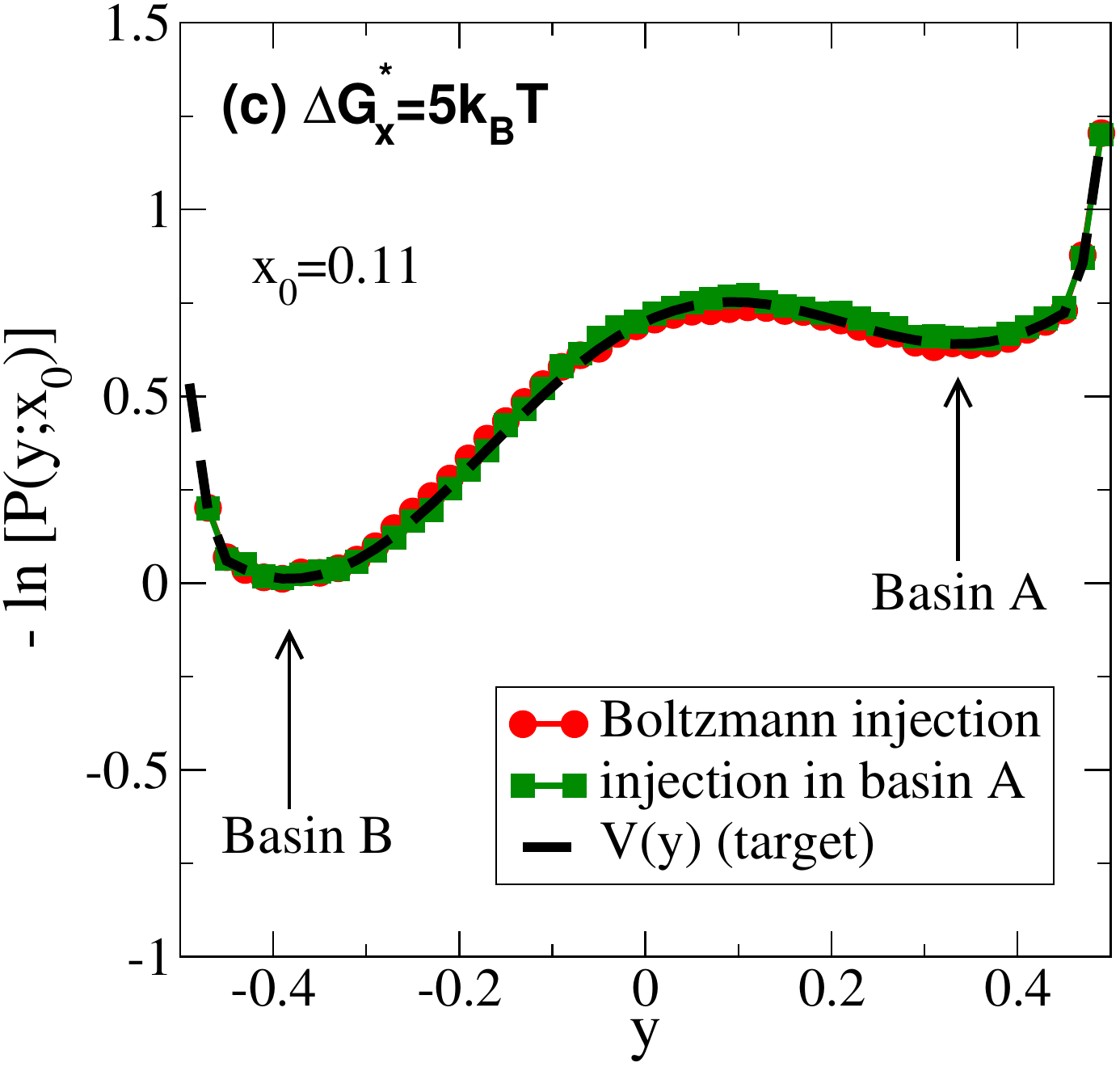}
\caption{The free energy reconstruction tested at $T=0.08$ for different barrier heights for $N_{traj}=600$ random walkers with a fixed maximum step size of $0.004$. In panel ({\bf a}) the reconstruction along $x$ is shown for two types of injection protocols as described in the legend. For the point injection protocol, $x=0$, $y=0.35$ in the $A$ basin is chosen. The results are compared with the kinetic reconstruction using the MFPT and against the effective barrier along $x$, $V_p(x)$ in Eq.~\ref{eq:vpx}. In panel ({\bf}) and ({\bf c}) a slice is taken along $x_0=0.11$ to compare with the ``target" slice $V(y;x_0)$ for different barrier heights along $x$ ($\Delta G(x^*)=1~k_BT$ in panel ({\bf b}) and $\Delta G(x^*)=5~k_BT$ in panel ({\bf c})). We find that for both the reconstruction along $x$ and along both $x$ and $y$, accurate reconstructions are possible regardless of the distribution of initial $y$ values.}
\label{fig:works_2D_1D}
\end{figure*}

\begin{figure*}[htpb!]
\centering
\includegraphics[scale=0.6]{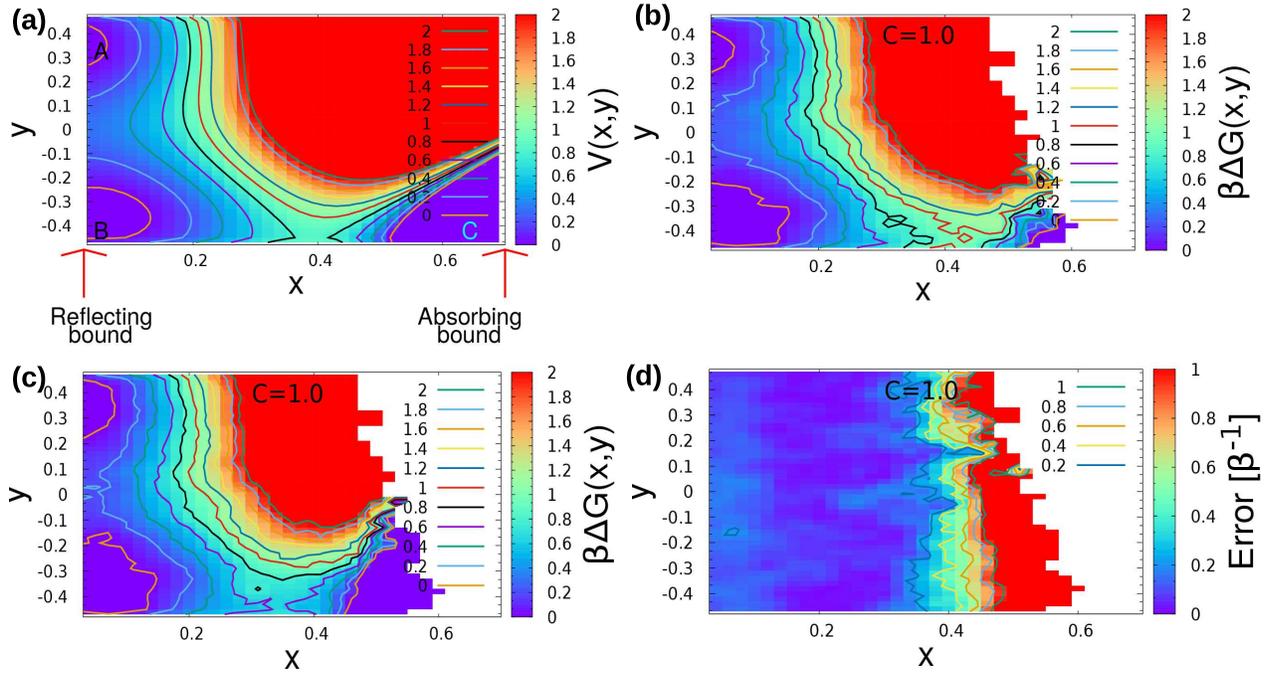}
\caption{
% \YGcom{
The full two-order parameter reconstruction for a case with a scaled barrier $1~k_BT$ barrier along $x$ (see Appendix~\ref{appsec:potential} for details on controlling  barrier height), as shown in the original potential energy surface in panel ({\bf a}. $N_{traj}=600$ random walkers are injected at $x=0$, $y=0.35$ and evolved with a temperature of $T=0.08$, maximum step size of $0.004$ and histograms binned with a square bin size of $0.02$. Panel ({\bf b}) shows the reconstruction of $\beta\Delta G(x,y)$ using Eq.~\ref{eq:bdgxy_hatpi} while panel ({\bf c}) shows a reconstruction using Eq.~\ref{eq:ratio_xy_so} $\beta \Delta G(x,y)$. 
% \sscom{I don't understand this. What are the two cases? What does C = 1 mean?}
The error between the reconstructed free energy and the supplied potential energy surface is measured as $|\beta\Delta G(x,y) - \beta V(x,y)|$ and shown in units of $k_BT$ in panel ({\bf d}). The errors are of the order of $<0.1~k_BT$ for $x<0.4$, which is approximately the value of $x^*$.
% }
}
\label{fig:canon_compare}
\end{figure*}

Writing 
\begin{equation}
\frac{P_{eq}(x,y)}{P_{eq}(A)}=\frac{\Phi_{A\rightarrow xy}}{\Phi_{xy\rightarrow A}}\frac{P_{st}(x,y)}{P_{st}(A)},
\end{equation} one obtains a correction factor that works for either a Boltzmann initialisation or a point injection at some $y_0$, $x=0$. In Fig. \ref{fig:works_2D_1D}, we show the reconstructed (single order parameter) free energy surfaces along $x$ and along $y$. We show the full free energy surface reconstruction and a comparison with errors in Fig.~\ref{fig:canon_compare}.

%\begin{equation}
%e^{- \beta V_p(x)} = \int_{-\infty}^{\infty}e^{-\beta V(x,y)}dy
%\label{eq:vpx}
%\end{equation}
The reconstruction of the single and two order parameter free energies using both Eq.~\ref{eq:ratio_xy_so} (Fig.~\ref{fig:works_2D_1D} (a) and Fig.~\ref{fig:canon_compare} (b)) and Eq.~\ref{eq:bdgxy_hatpi} (Fig.~\ref{fig:works_2D_1D} (a), (b), (c) and Fig.~\ref{fig:canon_compare} (c) , (d)) demonstrate the utility of this approach. Estimates of the free energies of the metastable states and the barrier heights agree quantitatively with the reference landscape.
% \YGcom{
The errors in Fig.~\ref{fig:canon_compare} (d) are $<0.1~k_BT$ for all $(x,y)$ in the metastable regime and are thus of the order of $1\%$.
% }
% \sscom{Quality this properly. 0.1 kbT seems pretty bad. }

The high errors beyond the barrier along $x$ arise from poor sampling of the backward flux back to the region defined as source ($A$ at $x=0$, $y=0.35$ or $x=0$ and all $y$ depending on injection protocol). In Section~\ref{sec:improving_estimates} we describe approaches to improve estimates beyond the barrier. 

We also reconstruct different free energy surfaces, having either more metastable states or multiple possible reaction pathways, to test the generality of our scheme. Results for these alternate potential surfaces are shown in Appendix~\ref{appsec:alt_potentials}. We next describe results for the reconstruction of the free energy landscape of supercooled silicon from unconstrained molecular dynamics simulations using this methodology.

%%%%%%%%%%%%%%%%%%%%%%%%%%%%%%%%%%%%%%%%%%%%%%%%%%%%%%%%%%%%%%%%%%%%%%%
%\FloatBarrier
\section{Results for supercooled liquid silicon}\label{sec:silicon_results}
% \YGcom{
We apply the methodology described above to the case of liquid silicon using the size of the largest crystalline cluster, $n_{max}$ and the density, $\rho$, as the order parameters with respect to which we reconstruct the free energy.
$n_{max}$ is analogous to $x$ in the test system and $\rho$ thus corresponds to $y$.
% }
% \sscom{You have not defined nmax.}
For each trajectory, labelled superscript $i$, we update the steady state sampling frequency of the values of $n_{max}$ and $\rho$ sampled by it using $N^i_{st}(n_{max},\rho)=\sum\limits_{t=0}^{t^i_{final}} \delta(n_{max}^i(t) - n_{max})\delta(\rho^i(t) - \rho)$. The steady state sampling probability $P_{st}(n_{max},\rho)$ is obtained by explicitly normalising with the sum of $N^i_{st}(n_{max},\rho)$ over all $n_{max}$ and $\rho$. 

The flux count is measured in the following way. First, we define the injection point of $n_{max}~<~1$ and $2.45~\leq~\rho~<2.46$ as the ``source" or reference state $A$. For each trajectory, at time step $t$, we consider the $n_{max}$ and $\rho$ values. If these are outside the region $A$, we trace back along the trajectory to check if $A$ was visited before this point. If it was visited (strictly always true for each $(n_{max},\rho)$), we update the count of the flux from $A$ to the given $(n_{max},\rho)$ by $1$. It must be ensured that multiple crossings from $A$ to a given $(n_{max},\rho)$ are not counted multiply. Likewise, at every time step at which the region $A$ is reached, we trace back along the trajectory and update the flux count from every $(n_{max},\rho)$ that was visited prior to the given timestep and which was not counted already.

\begin{figure*}[htpb!]
\includegraphics[scale=0.4]{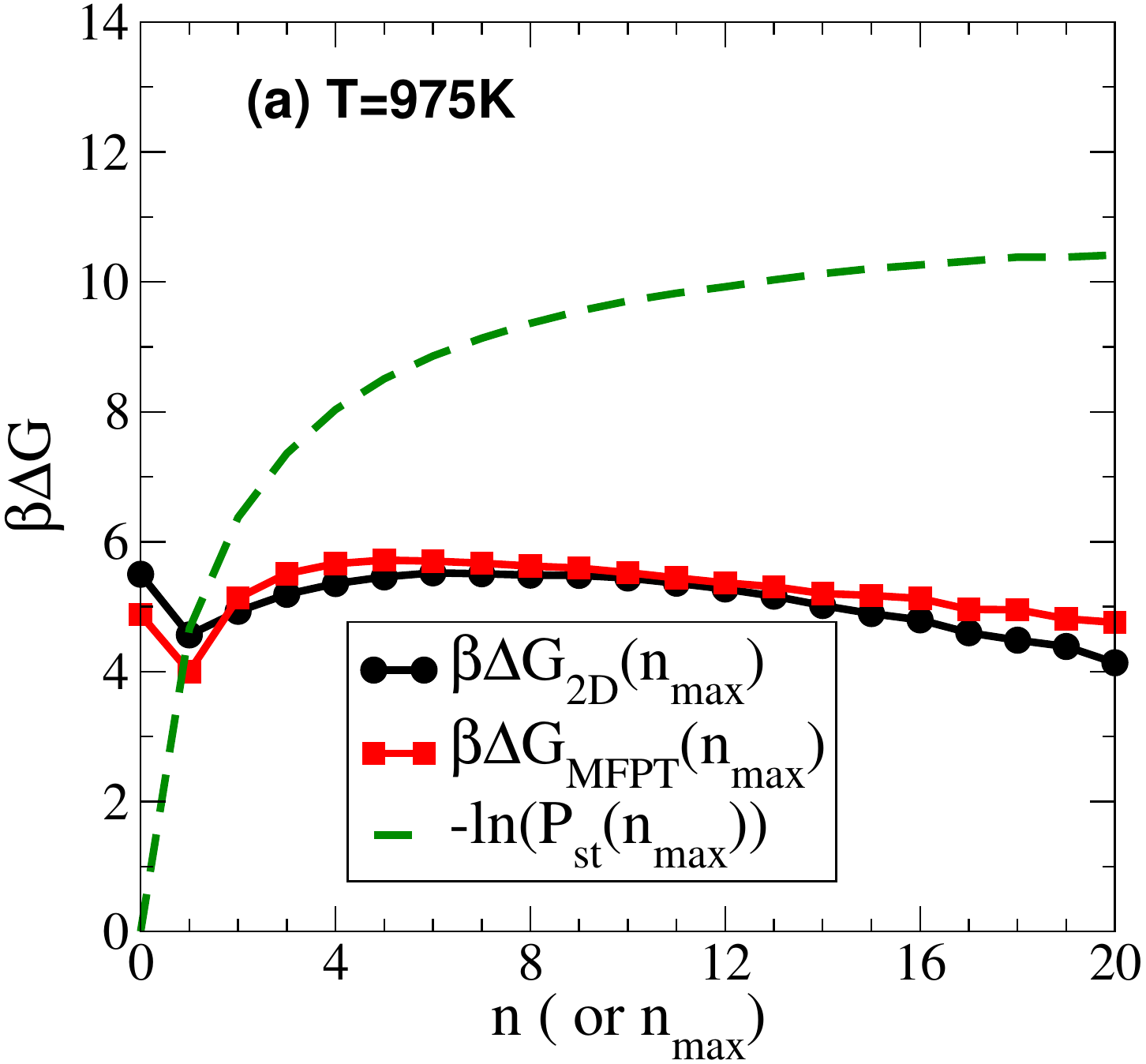}
\includegraphics[scale=0.4]{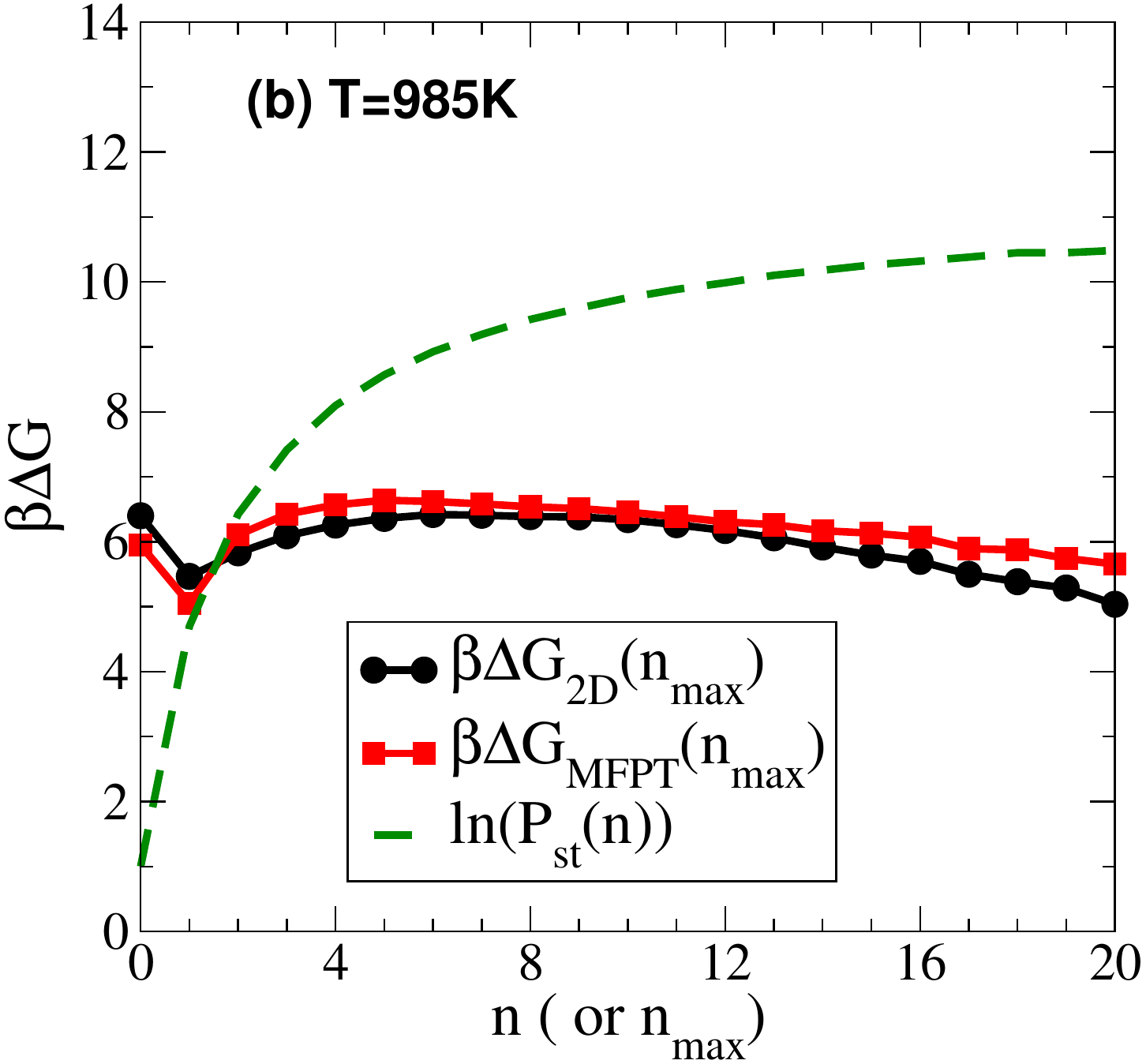}
\includegraphics[scale=0.4]{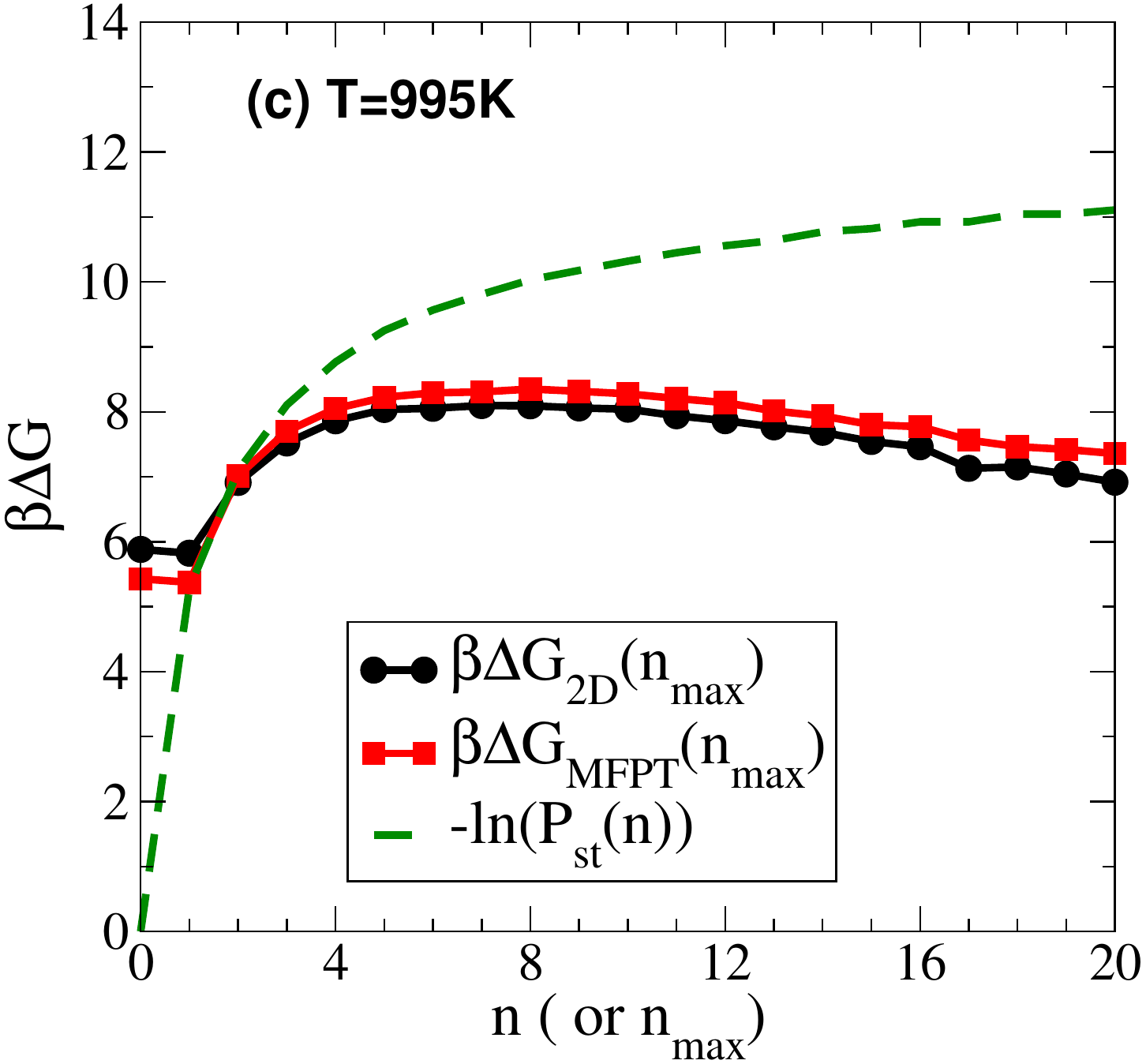}
\caption{The free energy as a function of largest crystalline cluster size ($n_{max}$) obtained using unconstrained MD simulations and reconstructed with either the mean first passage time (curves labelled MFPT) or Eq.~\ref{eq:ratio_xy_so} used for a single order parameter, $n_{max}$, (curves labelled 2D) for $T=975~K$, $P=0.75~GPa$ (panel {\bf a}), $T=985~K$, $P=0.75~GPa$ (panel {\bf b}) and $T=995~K$, $P=0.75~GPa$ (panel {\bf c}). In order to avoid artefacts due to the use of the largest cluster size as the order parameter, the curves are shifted to match the full cluster size distribution for $n_{max}~\leq~2$ as described in Ref.~\onlinecite{goswami2021thermodynamics}. The negative logarithm of the steady state full cluster size distribution is shown for reference in each case.}
\label{fig:si_recon_nmax}
\end{figure*}

We first compare estimates $\beta\Delta G(n_{max})$ obtained using a single order parameter version of Eq.~\ref{eq:ratio_xy_so} with those obtained from the MFPT method in Fig.~\ref{fig:si_recon_nmax}.  The small $n_{max}$ free energies are obtained by matching $\beta\Delta G(n_{max})$ with $-ln(P_{st}(n))$ from unconstrained MD runs for small $n$ (or   $n_{max}$)\cite{goswami2021thermodynamics}. $\beta\Delta G(n_{max})$ shows an artificial minimum, which is rectified from this comparison, as detailed in \cite{goswami2021thermodynamics}. We note that the results from the present method compare rather well with those of the MFPT method. 

\begin{figure*}[htpb!]
\includegraphics[scale=0.4]{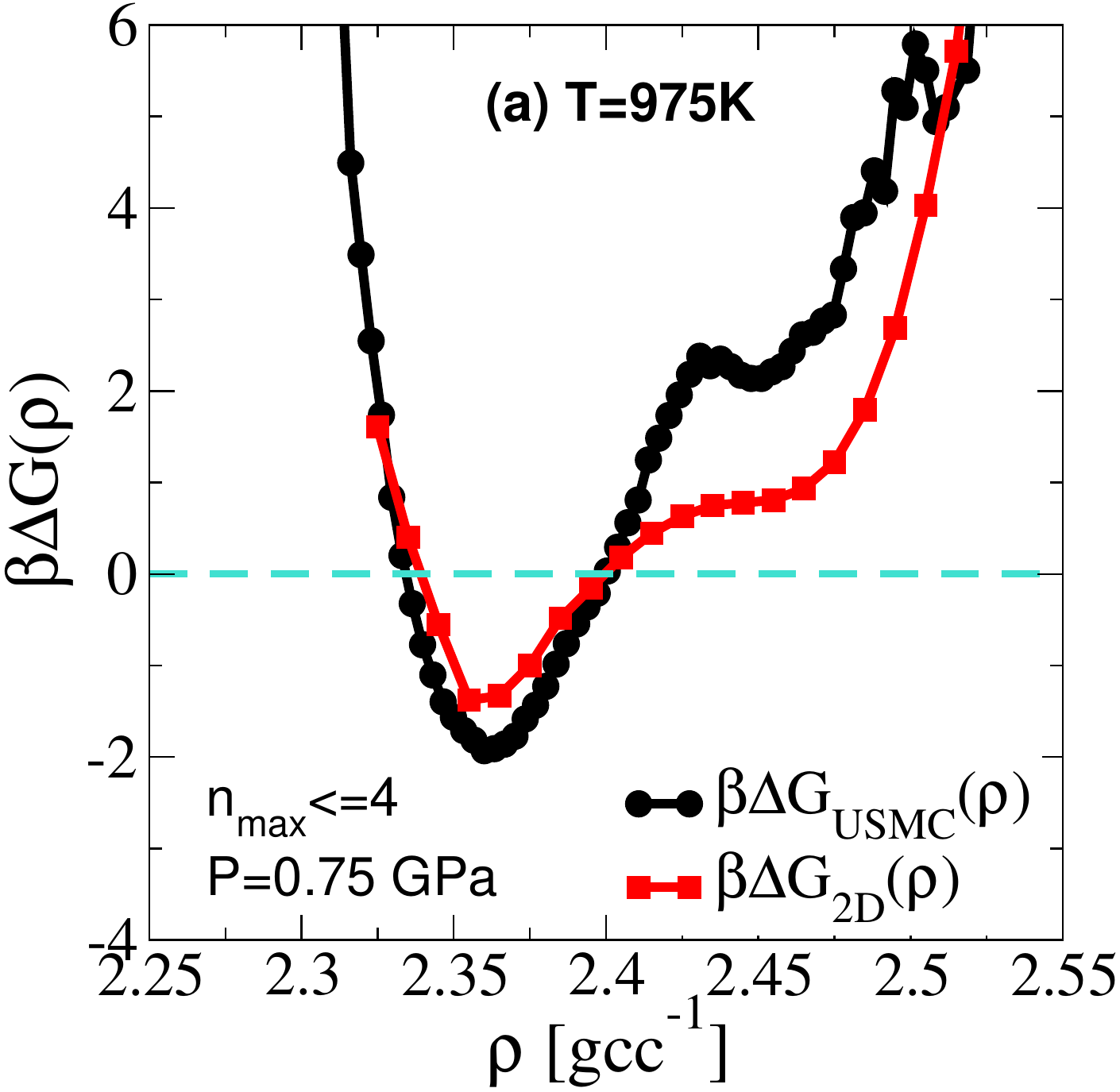}
\includegraphics[scale=0.4]{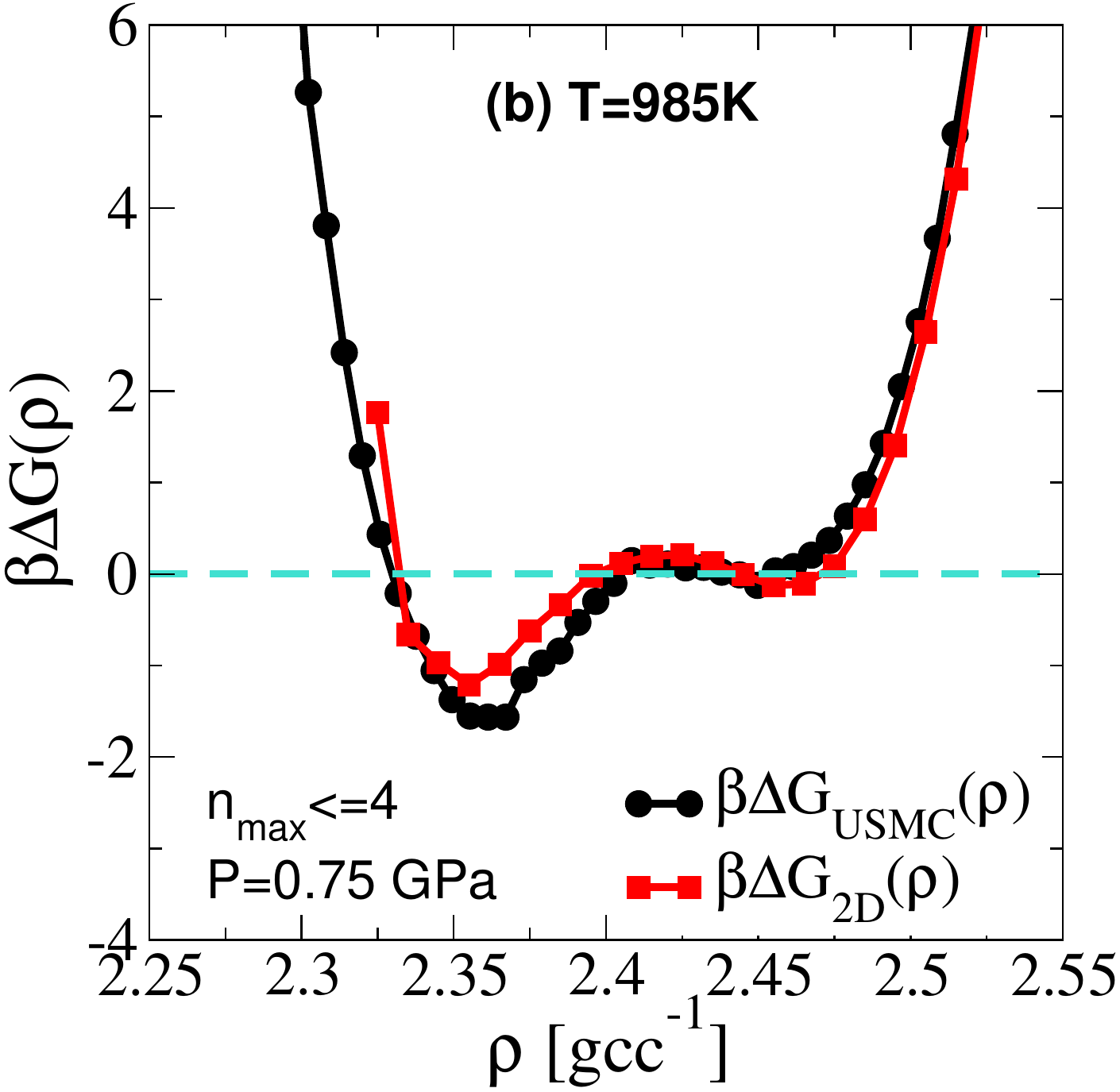}
\includegraphics[scale=0.4]{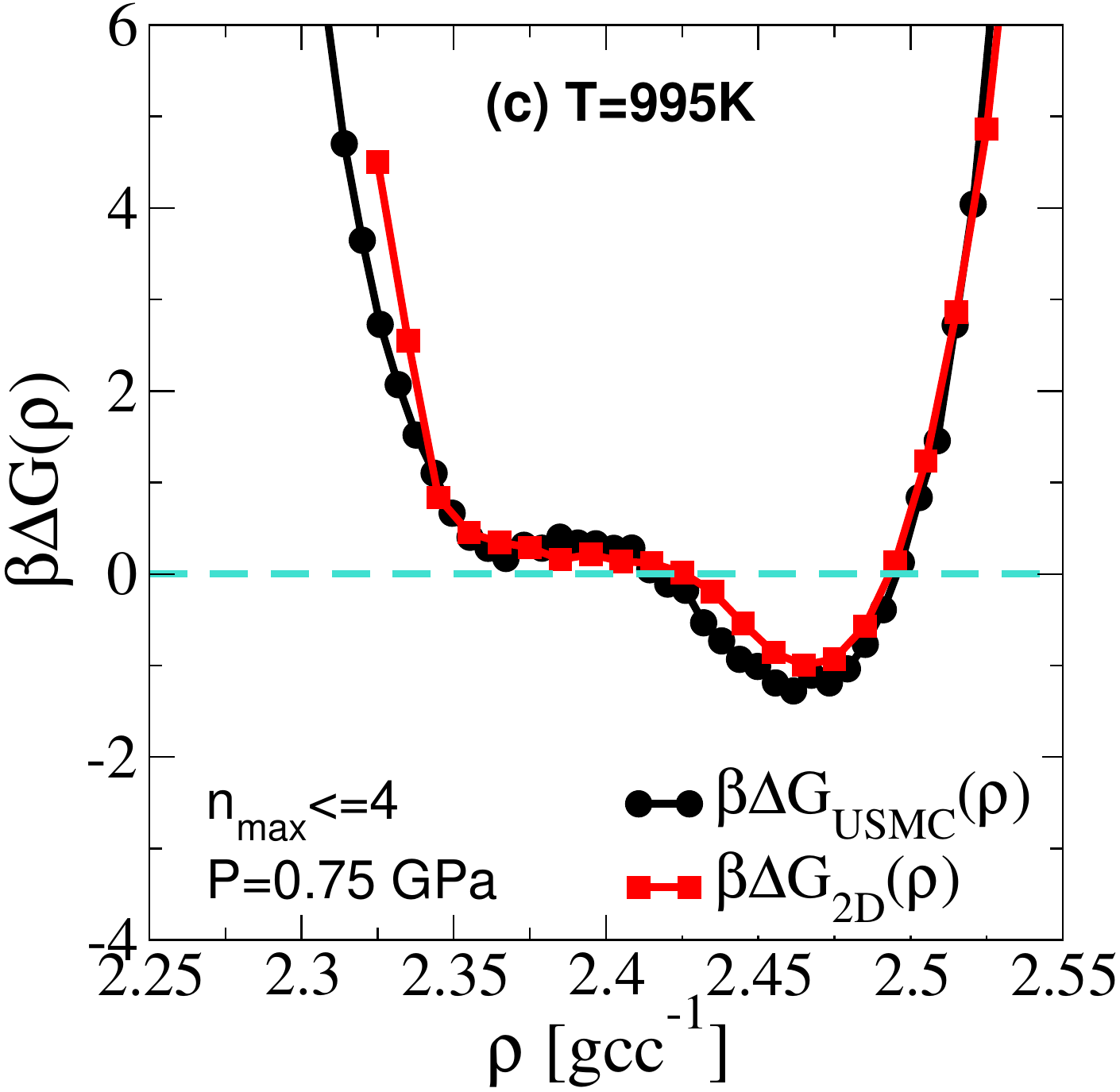}
\caption{The free energy as a function of density alone, from the region of order parameter space where the largest cluster size is less than the critical cluster size, reconstructed using either umbrella sampling Monte Carlo simulations (labelled USMC),
% \YGcom{
reprinted with permission from Goswami and Sastry, PNAS Nexus, \bf{$1$}, $4$, (2022). Copyright 2022 Author(s), licensed under a Creative Commons Attribution (CC BY) license.
% }
or Eq.~\ref{eq:bdgxy_hatpi} (labelled 2D) for $T=975~K$, $P=0.75~GPa$ (panel {\bf a}), $T=985~K$, $P=0.75~GPa$ (panel {\bf b}) and $T=995~K$, $P=0.75~GPa$ (panel {\bf c}). The free energy as a function of density is obtained from the full two-order parameter distribution using Eq.~\ref{eq:Prho_conditional}.}
\label{fig:si_recon_rho}
\end{figure*}

We obtain the free energy in terms of the density $\rho$, $\beta \Delta G(\rho)$, upto an irrelevant additive constant from the full probability distribution $P(n_{max},\rho)$ by using
\begin{equation}
   P(\rho)=\sum\limits_{n_{max}=0}^{n_{max}=4}P(n_{max},\rho)
   \label{eq:Prho_conditional}
\end{equation} 
and taking the negative logarithm. 
% The results are quantitatively compared with the free energy as a function of density obtained from umbrella sampling simulations reported in\cite{goswami2022liquid} 
In Fig.~\ref{fig:si_recon_rho} the reconstructed free energy as a function of density, $\beta\Delta G(\rho)$ is obtained by using Eq.~\ref{eq:bdgxy_hatpi} to reconstruct the free energy surface and Eq.~\ref{eq:Prho_conditional} to get $\beta\Delta G(\rho)$. 
These results are compared with corresponding results from umbrella sampling runs constraining both $\rho$ and $n_{max}$ for which the data is obtained from Ref.~\onlinecite{goswami2022liquid}.
The density profiles show a shift in the location of the metastable minimum in density from a high value of $2.45~gcc^{-1}$ to a low density of $2.35~gcc^{-1}$ when the temperature is changed from $T=995~K$ to $T=985~K$ at $P=0.75~GPa$ with the other liquid state losing metastability at or around $T=985~K$. At $T=975~K$, one observes a larger difference in the estimates for $\beta\Delta G(\rho)$ obtained using the two methods. This remains to be fully understood, possible reasons being poor sampling of the high density liquid in the case of umbrella sampling, given the high local variation observed, or limitations arising from the assumption of invariant rate in writing Eq.~\ref{eq:Pst_to_Peq_orig}.

In the next section, we discuss possible ways to improve on the methodology, addressing the shortcomings of poor estimates beyond the barrier and the possible sources of discrepancy in the results for silicon.
\section{Strategies for improved free energy estimates}\label{sec:improving_estimates}
Given states $A$ and $B$ between which we want to measure the flux, $\langle \Phi_{AB} \rangle$ and $\langle \Phi_{BA} \rangle$, the quality of reconstruction is determined both by the extent of sampling the steady state probability as well as the two fluxes. In this section we discuss approaches to improve the sampling of order parameter space and therefore the resulting free energy estimates by addressing these requirements.
\subsection*{Using interfaces for accurate flux calulation}\label{subsec:interface}
\begin{figure*}[htpb!]
\centering
\includegraphics[scale=0.6]{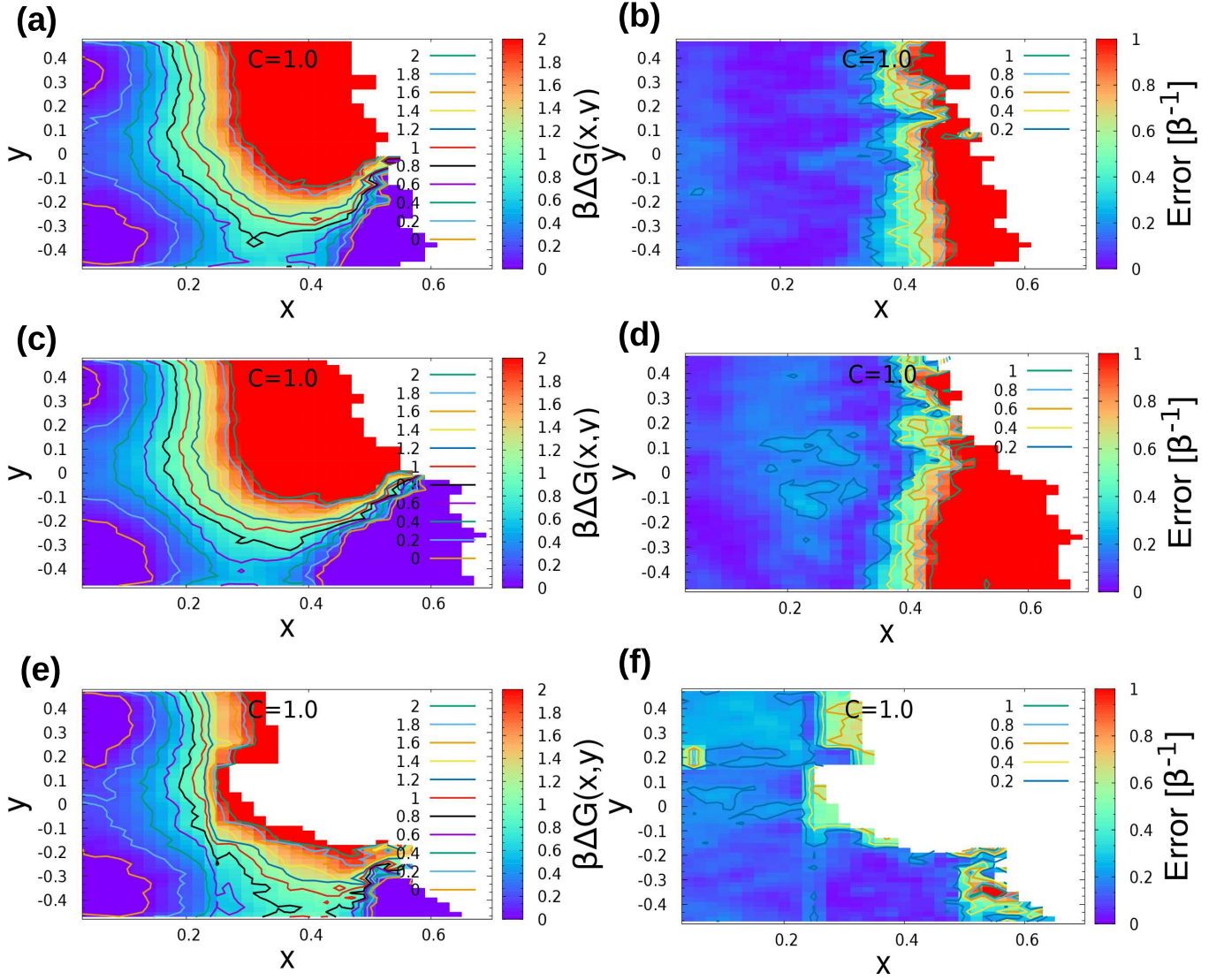}
\caption{A comparison of the the free energy surface, $\beta\Delta G(x,y)$ using Eq.~\ref{eq:bdgxy_hatpi} (panel ({
\bf a}) and error $|\beta\Delta G(x,y) - \beta V(x,y)|$ from the same procedure in units of $k_BT^{-1}$ (panel ({\bf b}). The corresponding free energy surface and errors using Eq.~\ref{eq:int_improve_estimates} with an interface placed at $x=0.5$ are shown in panels ({\bf c}) and ({\bf d}) respectively. One observes an improvement in estimates beyond the barrier due to the improved estimates of the backward flux. Data is masked for $9x,y)$ values where errors exceeding the scale shown. Panels ({\bf e}) and ({\bf f}) show results using the procedure of combining estimates from multiple sections with unique reference states $A$.}
\label{fig:2OPMDB_improved_estimates}
\end{figure*}

We first describe how to improve the sampling of the backward flux from $B$ to $A$ for regions $B$ that are beyond the barrier along $x$, such that this backward flux is low and therefore poorly sampled with a finite number of trajectories. Better estimates of this flux can be obtained by placing an interface between $A$ and $B$ and expressing the total flux as a product. This approach can be used if and only if every trajectory from $A$ to $B$ and $B$ to $A$ passes through an intermediate, $I$, (different from the new absorbing condition $C$). We can then write the following
\begin{align}
\Phi_{A\rightarrow B} &= \Phi_{A\rightarrow I}\times \Phi_{I\rightarrow B} \nonumber \\
\Phi_{B\rightarrow A} &= \Phi_{B\rightarrow I}\times \Phi_{I\rightarrow A} \nonumber
\end{align}
% If this condition is not satisfied, the expression is more complicated, involving seggregating trajectories that go from $A$ to $B$ (or $B$ to $A$) without going through $I$ first.
Choosing $I$ as a hyper-plane separating $A$ and $B$ ensures that this condition is met. In the 2D case, $I$ is a line. 
% \sout{In cases of higher dimension, and when the order parameter is not known, defining $I$ is harder to do.}
We proceed by testing if estimates can be improved for $x$ beyond the barrier, from where the flux back to $A$ may be negligibly small.
We place the line at $x=0.5$, beyond the saddle. As a test, we can compare and check if the following equation is true
\begin{equation}
\phi_{A\rightarrow xy}=\phi_{A\rightarrow I} \times \phi_{I\rightarrow xy},
\label{eq:int_improve_estimates}
\end{equation}  
for each $(x,y)$ with $x>0.5$. This is easy to verify because there is a large direct flux from $A$ to $(x,y)$ beyond the barrier.
% We find that it is true through a recalculation of the free energy surface using either $\phi_{so\rightarrow xy}$ or $\phi_{so\rightarrow I} \times \phi_{I\rightarrow xy}$ (see Fig.~\ref{fig:2OPMDB_improved_estimates}). We also find that estimates of free energy for $x>0.5$ are improved as a result.
We find that the free energy reconstruction is improved beyond $x=0.5$ by using this expression for the flux. Fig.~\ref{fig:2OPMDB_improved_estimates} (c) and (d) show the results from this procedure and can be compared with panels (a) and (b) respectively in Fig.~\ref{fig:2OPMDB_improved_estimates}, which are obtained without resolving the flux along the lines in Eq.~\ref{eq:int_improve_estimates}.
It should be noted that for $(x,y)$ beyond the barrier and close to the absorbing state $C$, our assumption that typical paths from $A$ to $xy$ and the reverse do not pass through the neighbourhood of $C$ does not hold. As a consequence, the phenomenological rate changes significantly when one or both of $A$ and $B$ are close to $C$, possibly also lacking a timescale over which its value is a plateau.
% The requirement of a plateau in the time-dependent value of the rate in Eq.~\ref{eq:TPS_rate_defn_app} for $\tau_{mol} < t \ll t_{ArB}^{sp}$ in the presence of $C$ does not hold.

\subsection*{Sectioned reconstruction with different reference states}\label{subsec:sections}
We next attempt to improve the free energy estimates by defining multiple sections of the order parameter space, each with a unique reference $A$ state. The equilibrium probability $P_{eq}(x,y)$ is obtained using Eq.~\ref{eq:ratio_xy_so} independently in each section using the fluxes with respect to the unique reference state. We ensure that each reference state is also contained within another neighbouring region so that the estimates for each reference region can be matched with the estimate from the neighbouring region by a simple shifting of $-ln(P_{eq}(A))$. This then gives a better estimate of fluxes locally and the different free energy estimates are then shifted to reconstruct the full surface. In Fig.~\ref{fig:2OPMDB_improved_estimates} (e) and (f), we demonstrate the results of this procedure for a case where the order parameter space is divided into a $3\times3$ grid. The bottom left corner of each region is chosen as the reference state for it. The region sizes are larger than the separation between the reference points so that there is overlap. The errors, shown in panel (f), are significantly lower at large $x$ values than the corresponding errors in panels (b) and (d), where the reconstruction was performed using Eq.~\ref{eq:ratio_xy_so} and Eq.~\ref{eq:int_improve_estimates} respectively. 

% \sscom{Isn't the subsection below a repetition?}
% \YGcom{Removed repeated subsection on results with modified methods}
% \subsection*{Comparison of results with improved estimates}
% In Fig.~\ref{fig:2OPMDB_improved_estimates}, we show how an improved estimate of the backward flux can enhance sampling beyond the barrier along $x$ for the system of independent random walkers on a potential energy surface. The results using Eq.~\ref{eq:ratio_xy_so} and the corresponding error are shown in panels (a) and (b), while those using Eq.~\ref{eq:int_improve_estimates} are shown in panels (c) and (d). Finally, in panels (e) and (f), the free energy reconstruction and the corresponding error are shown by combining estimates from $9$ sections. 

\section{Discussion}\label{sec:discussion}
We have described an effective and efficient method to obtain estimates of free energies as a function of multiple order parameters from unconstrained simulations. 
With our motivation arising from the study of polymorphism in supercooled liquid silicon, we address the problem of reconstructing a multi-dimensional free energy surface that can distinguish the possible metastable states as well as the globally stable crystalline state.
In order for simulation lengths to be tractable, an absorbing condition is placed at large values of the crystallinity order parameter. This absorbing boundary introduces a flux in the system altering sampling. By treating the trajectories in the order parameter space as obeying Brownian motion in the high-friction limit one can obtain the steady state sampling in order parameter space, as well as the various point-to-point fluxes. We consider the expression derived for the transition rate, expressed in terms of sampling probability and fluxes, to determine a relation between steady state sampling in the presence of a non-zero constant flux (to the imposed absorbing boundary) to the underlying equilibrium sampling, which can otherwise only be achieved under conditions of zero net flux, which describes detailed balance. 
This relationship between the steady state sampling and the underlying equilibrium sampling is the key aspect of our method, which allows us to obtain the free energies without the added effort of ensuring the zero flux condition. We show that the method works efficiently for multiple cases by testing it on a test system of random walkers on a potential energy landscape. We choose potential energy landscapes that have, in addition to the globally stable state, (i) multiple metastable states, (ii) metastable states that are not on the primary ``transition tube" connecting the initial metastable state to the final, globally stable state (see Appendix~\ref{appsec:alt_potentials} Fig.~\ref{fig:3meta_compare}), and (iii), multiple paths to the globally stable state with multiple saddles of different heights connecting the metastable states to the globally stable state (see Appendix~\ref{appsec:alt_potentials} Fig.~\ref{fig:two_saddle_compare}). We find that the method reconstructs the free energy accurately and efficiently in the metastable region provides good estimates of basin depth and barrier height. Issues of poor sampling affect the calculation close to the absorbing boundary and the deep minimum of the globally stable state. We discuss and demonstrate multiple methods to improve our estimates.

This method is applied to unconstrained molecular dynamics trajectories of supercooled liquid silicon, for which free energy calculations recently performed using umbrella sampling Monte Carlo exhibited two metastable liquid states. Using the approach described here, we are able to reproduce these free energy estimates, confirming the applicability of such an approach in a more realistic and challenging context.

Quantitative differences in the results for silicon between the first passage time reconstruction and the method described here, as well as differences at the lowest temperature studied here remain to be understood better. Moreover, more exact treatments of the rate in the presence of perturbations away from equilibrium remain an important open issue to address. The connection to driven systems is also of importance\cite{crooks1998nonequilibrium,kuznets2021dissipation,das2022annealing}, though the role of the free energy is less clear for a system driven from equilibrium with an external field.

The framework of the population flux correlation function in defining phenomenological rates is a significant milestone in the study of rare events and in the subsequent development of free energy calculation methods\cite{chandler1978statistical}. Here, by considering the issues arising from a low barrier and thus a high flux, one can better address the relevant physics in regimes where a number of approximations common to the high barrier regime do not apply. Future work that addresses the limitations identified here and strengthens the connection to driven systems are important avenues to explore.


\FloatBarrier
% \begin{acknowledgments}
% % The authors gratefully acknowledge the Thematic Unit of Excellence on Computational Materials Science, and the National Supercomputing Mission facility (Param Yukti) at the Jawaharlal Nehru Center for Advanced Scientific Research for computational resources.  SS acknowledges support through the JC Bose Fellowship  (JBR/2020/000015) SERB, DST (India). The authors gratefully acknowledge Mustansir Barma, Daan Frenkel,  Meghna A. Manae and Muhittin Mungan for helpful discussions.
% \end{acknowledgments}

\section*{Data Availability Statement}
The data that support the findings of this study are available from the corresponding author upon reasonable request.
\FloatBarrier
\appendix
\section{Potential details}\label{appsec:potential}
The values of the means and standard deviations for the Gaussian components of the potential in Fig.~\ref{fig:vxy_form} are shown in Table~\ref{table:pot_details}. 
% \begin{align*}
% V_1(x,y) &= -C_1\left( e^{\left [\frac{(x-x_1)^2}{\sigma x_1} + \frac{(y-y_1)^2}{\sigma y_1}\right ]} \right ) \nonumber \\
% V_2(x,y) &= -C_2\left( e^{\left [\frac{(x-x_2)^2}{\sigma x_2} + \frac{(y-y_1)^2}{\sigma y_2}\right ]} \right ) \nonumber \\
% V_3(x,y) &= -C_3\left( e^{\left [\frac{(x-x_3)^2}{\sigma x_3} + \frac{(y-y_3)^2}{\sigma y_3}\right ]} \right ) \nonumber \\
% V_4(x,y) &= +C_4\left( e^{\left [\frac{(x-x_4)^2}{\sigma x_4} + \frac{(y-y_4)^2}{\sigma y_4}\right ]} \right ) \nonumber \\
% \end{align*}
\begin{table}[h]
\centering
\begin{tabular}{|l|l|l|l|l|l|}
\hline
	$V_i$ & $x_i$ & $y_i$ & $\sigma x_i$ & $\sigma y_i$ & $C_i$\\
	\hline
	$V_1$ & 0.05 & -0.4 & 0.2 & 0.2 & -0.2\\
	$V_2$ & 0.01 & 0.4 & 0.2 & 0.2 & -0.2\\
	$V_3$ & 1.0 & -0.4 & 0.1 & 0.1 & -2.0\\
	$V_4$ & 0.7 & 0.4 & 0.08 & 0.08 & C\\
\hline
\end{tabular}
\caption{The table of factors used to specify the potential energy surface defined in Eq.~\ref{eq:vxy}. The value of $C$ is modulated to produce barriers along $x$ ranging from $0.08$ for $C=1$ to $0.16$ for $C=4$.}
\label{table:pot_details}
\end{table}
The harmonic potential to ensure sampling remains within $-[0.5,0.5]$ is specified as  $V_c(y) = k_c(|y| - 0.45)^2$ with a $k_c$ value of $20$.

\section{Tests on an alternate potential - 3 metastable basins or two saddles}\label{appsec:alt_potentials}
\begin{figure*}[htpb!]
\centering
\includegraphics[scale=0.6]{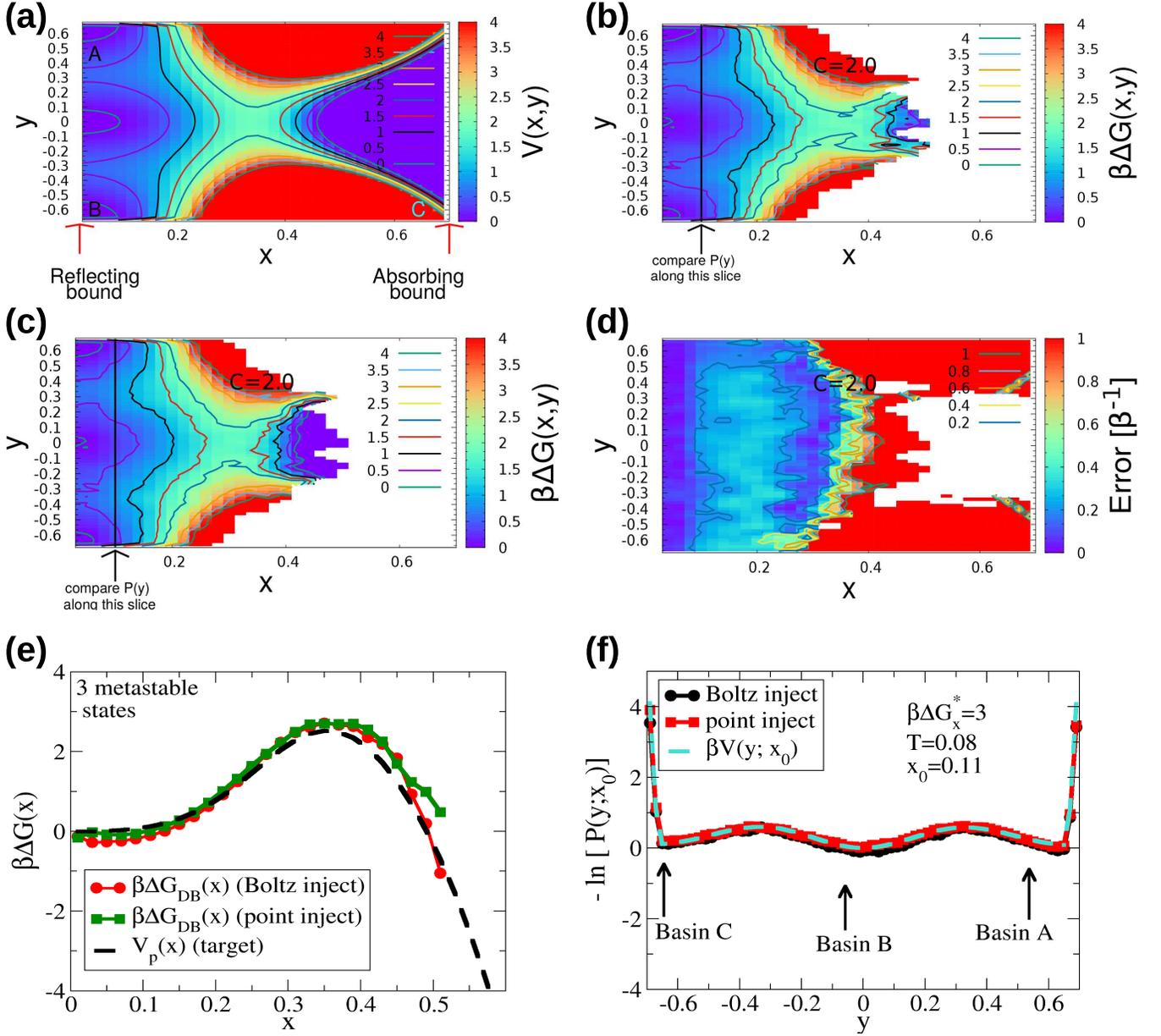}
\caption{Potential surface with three metastable basins and one globally stable state (panel ({\bf a})).
Reconstruction of the two order parameter free energy using Eq.~\ref{eq:ratio_xy_so} (panel ({\bf b})). Reconstruction of the two order parameter free energy using Eq.~\ref{eq:bdgxy_hatpi} (panel ({\bf c})).
The relative error, $|\beta\Delta G(x,y) - \beta V(x,y)|$, units of $k_BT$, from the reconstruction using Eq.~\ref{eq:bdgxy_hatpi} (panel ({\bf d})).
Reconstructions of the one order parameter free energy along $x$ (panel ({\bf e})) and for a slice along $y$ (panel ({\bf f})).}
\label{fig:3meta_compare}
\end{figure*}

A potential with 3 meta-stable basins, an effective barrier height along $x$ of $3~k_BT$  and one basin, C, not part of the reaction path (assuming injection at A) to the globally stable D (see Fig.~\ref{fig:3meta_compare} (panel (a)). Panels (b) and (c) show the reconstruction using Eq.~\ref{eq:ratio_xy_so} and Eq.~\ref{eq:bdgxy_hatpi} respectively. Errors are shown in Fig.~\ref{fig:3meta_compare} (d) to show the degree of accuracy. Fig.~\ref{fig:3meta_compare} shows the reconstruction of the free energy surface and the contracted free energy along $x$ (panel (e)) and the comparison for a slice along $y$ (panel (f)).

\begin{figure*}[htpb!]
\centering
\includegraphics[scale=0.6]{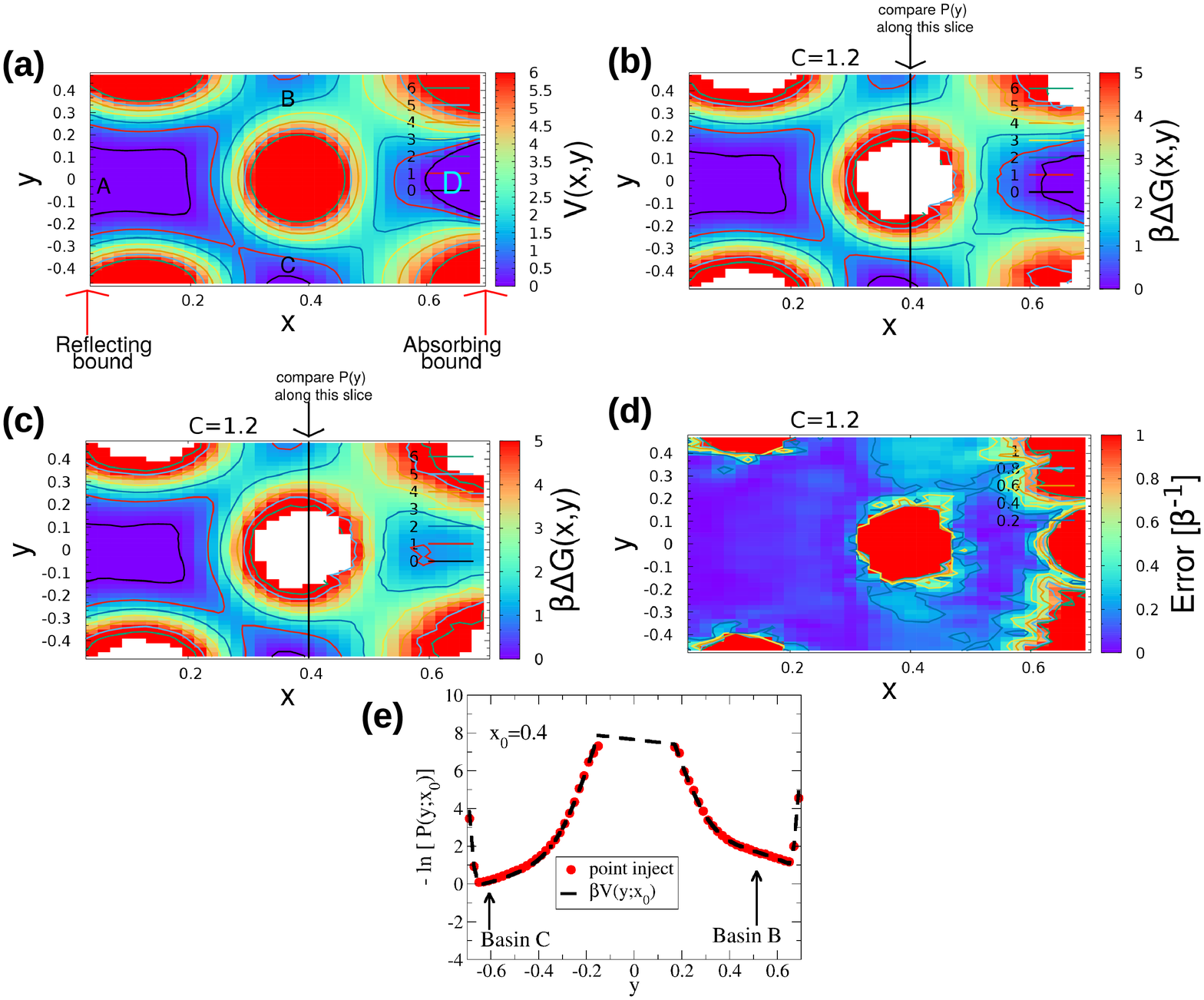}
\caption{A potential surface with 2 saddles separating metastable states, $A,B,C$ from global minimum, $D$ (panel ({\bf a})). Reconstruction of the two order-parameter free energy using Eq.~\ref{eq:ratio_xy_so} (panel ({\bf b})). Reconstruction of the two order parameter free energy using Eq.~\ref{eq:bdgxy_hatpi} (panel ({\bf c})). 
The relative error, $|\beta\Delta G(x,y) - \beta V(x,y)|$, units of $k_BT$ from a reconstruction using Eq.~\ref{eq:bdgxy_hatpi} (panel ({\bf d})). Comparison of the free energy for a slice along $y$ for a fixed $x$ (panel ({\bf e})).}
\label{fig:two_saddle_compare}
\end{figure*}

%\FloatBarrier
A potential with two paths separating metastable basins from the globally stable basin. The heights of the saddles along the two paths are unequal to introduce an asymmetry (see Fig.~\ref{fig:two_saddle_compare}). Fig.~\ref{fig:two_saddle_compare} shows the reconstruction of the free energy surface and the contracted free energy along $x$. Errors are shown in Fig.~\ref{fig:two_saddle_compare} to show the degree of accuracy.
The results in this section show that the method to reconstruct free energies is robust to free energy landscapes with a variety of features.

\FloatBarrier

%%%%%%%%%%%%%%%%%%%%%%%%%%%%%%%%%%%%%%%%%%%

% \section{A little more on appendixes}

% Observe that this appendix was started by using
% \begin{verbatim}
% \section{A little more on appendixes}
% \end{verbatim}

% Note the equation number in an appendix:
% \begin{equation}
% E=mc^2.
% \end{equation}

% \subsection{\label{app:subsec}A subsection in an appendix}

% You can use a subsection or subsubsection in an appendix. Note the
% numbering: we are now in Appendix~\ref{app:subsec}.

% \subsubsection{\label{app:subsubsec}A subsubsection in an appendix}
% Note the equation numbers in this appendix, produced with the
% subequations environment:
% \begin{subequations}
% \begin{eqnarray}
% E&=&mc, \label{appa}
% \\
% E&=&mc^2, \label{appb}
% \\
% E&\agt& mc^3. \label{appc}
% \end{eqnarray}
% \end{subequations}
% They turn out to be Eqs.~(\ref{appa}), (\ref{appb}), and (\ref{appc}).

\nocite{*}
\bibliography{2OP_reconstruction}% Produces the bibliography via BibTeX.

\end{document}